\documentclass[12pt]{article}

\usepackage[a4paper,text={15cm,23cm},centering]{geometry}
\usepackage{latexsym}
\usepackage{amsmath}
\usepackage{amssymb}
\usepackage{graphicx}
\usepackage{mathrsfs}
\usepackage{eucal}
\usepackage{bm}
\usepackage{epsfig}
\usepackage{subcaption}
\usepackage{wrapfig}
\usepackage{color}

   \graphicspath{{./figures/}}

\usepackage[bookmarks=false]{hyperref}


\font\bbf=cmbx12 scaled\magstep2

\def\V{{\mathbb V}}

\def\eps{\varepsilon}

\def\lth{\langle\!\langle}
\def\rth{\rangle\!\rangle}

\def\Zh{\widehat Z}
\def\eps{{\varepsilon}}

\def\lth{\langle\!\langle}
\def\rth{\rangle\!\rangle}

\def\D{\mbox{${\rm D}$}}

\def\bw{{\bm \omega}}
\def\bq{{\bf q}}
\def\bk{{\bf k}}
\def\bO{{\bm \Omega}}
\def\be{{\bm \zeta}}
\def\bth{{\bm \theta}}
\def\bph{{\bm\phi}}

\newcommand{\red}[1]{\textcolor{red}{#1}}

\begin{document}

\begin{center}
{\bbf Phase Dynamics of the Dysthe equation and the Bifurcation of Plane Waves}\\
\vspace{.5cm}
D.J. Ratliff\\
\vspace{3mm}
\emph{Loughborough University, Epinal Way, Leicestershire, U.K., LE11 3TU.}
\end{center}
\vspace{.5cm}

\begin{abstract}
The bifurcation of plane waves to localised structures is investigated in the Dysthe equation, which incorporates the effects of mean flow and wave steepening. Through the use of phase modulation techniques, it is demonstrated that such occurrences may be described using a Korteweg - de Vries (KdV) equation. The solitary wave solutions of this system form a qualitative prototype for the bifurcating dynamics, and the role of mean flow and steepening is then made clear through how they enter the amplitude and width of these solitary waves. Additionally, higher order phase dynamics are investigated, leading to increased nonlinear regimes which in turn have a more profound impact on how the plane waves transform under defects in the phase.
\end{abstract}

\vspace{5mm}

\vspace{5mm}
\section{Introduction}

At the heart of the modern study of waves is their behaviour and stability. The last century has heralded many studies and successes into these avenues, but there is much left to be understood. Particularly, the stability of monochromatic wavetrains in hydrodynamics generated large interest after the experiments of Benjamin and Feir demonstrated that such a state was unstable in experiments \cite{bf67,b67}; some analytical insight has since been gained using various mathematical techniques \cite{sd78}. Such instabilities have been speculated to lead to the formation of rogue waves \cite{oos00,srk07}, or the decrease in the frequency of the monochromatic wave \cite{lyr77,bhl86,shc05}. However, even when such waves are stable to an instability like this, they can undergo different transitions to generate structures such as dark solitary waves \cite{mvz13,i00,lzw17}, whose mechanism for formation remains unclear. Therefore, even such a heavily studied problem has a wealth of dynamics we have yet to understand.

The role of mean flow coupling on both the dynamics and stability of water waves has long been emphasised. There have been many contributions towards accurately providing a description between this interplay. Early work by Davey and Stewartson \cite{ds74} provided a system which coupled a Nonlinear Schr\"odinger equation to an irrotational fluid flow. This work has been furthered by several authors since. Of particular note is the work of Dysthe et al. \cite{d79,td97}, who have derived several higher order models which capture the effects of this mean flow-wave coupling with a large degree of success. For example, these have helped to more accurately investigate the stability of uniform wavetrains \cite{td96,td97,tkd00}, rogue wave formation \cite{kls04,sdk05} and frequency downshifting \cite{dtk03}.
It is because of this rich variety of behaviours that this paper will concern itself with the Dysthe equation with only its leading order dispersive term, as written in \cite{td97}. Thus, the governing equations are given by
\begin{equation}\label{DystheA}
\begin{split}
iA_t+A_{xx}-|A|^2A&+\alpha \left.\Phi_x\right\vert_{z=0}A+i\beta|A|^2A_x = 0\,,\\[2mm]
&\left.\Phi_z\right\vert_{z=0} = \alpha (|A|^2)_x\,,\\[2mm]
&\nabla^2\Phi = 0\quad \mbox{in} \quad z\in (-h,0)\,,\\[2mm]
&\left.\Phi_z\right\vert_{z = -h} = 0\,,
\end{split}
\end{equation}
for complex-valued wave envelope $A(x,t)$, velocity potential $\Phi(x,z,t)$ of the flow and $h$ the depth of the fluid. The constants $\alpha$ and $\beta$ characterise the magnitude of mean-flow and higher order self-steepening effects respectively. In the work of Trulsen and Dysthe \cite{td97}, the values of these parameters are taken to be $\alpha = 4$ and $\beta = 8\,,$ but we shall leave these free in order to see their explicit role in the emergence and evolution of defects.

In order to investigate systems which admit wavetrains and discuss their stability properties, one may utilise a phase dynamical approach. The origins of such studies can be traced back to Whitham \cite{w65,wlnlw}, with several years of subsequent work building upon these ideas. The premise is to assume that one has a wavetrain solution of the form $\hat{u}(kx+\omega t) \equiv \hat{u}(\theta;k,\omega)$ for phase $\theta$, wavenumber $k$ and frequency $\omega$. Then, the key idea is to assume that these wave variables are not fixed, but instead slowly vary in time. Under this assumption, either by an averaging principle or direct asymptotic analysis, one generates the system of equations
\begin{equation}\label{WME}
A(k,\omega)_T+B(k,\omega)_X = 0\,, \qquad k_T = \omega_X\,,
\end{equation}
for the now slowly varying wavenumber $k(X,T)$, frequency $\omega(X,T)$ and slow variables $X = \eps x,\,T = \eps t$ for $\eps \ll 1$. The functions $A$ and $B$ turn out to be the wave action and wave action flux respectively for the original system, averaged over one period of the original wavetrain $\hat{u}$. This set of equations then govern how the wavenumber and frequency evolve, which lead to deformations in the wavetrain from which they are derived. There have been several extensions to the Whitham methodology, such as the extension to problems with multiple phases \cite{ab70,r18,wlnlw} and to the more general setting of relative equilibria \cite{b17,w67,wlnlw}. It is these extensions that will allow us to explore the phase dynamics of the Dysthe system (\ref{DystheA}), as one not only modulates the emergent wavetrain but also the mean flow, which itself forms a relative equilibrium.

The Whitham equations (\ref{WME}) have the additional usage that, by investigating the linearised stability problem for a fixed wavenumber and frequency, one can infer stability properties of the original wave. This is diagnosed by the properties of the eigenvalues of such a linearisation, which are denoted as the characteristics of the system. For real characteristics, the system is hyperbolic and the associated wavetrain is stable. Alternatively, in cases where the characteristics become complex the Whitham system is elliptic and the underlying wave is unstable, with perturbations to it growing exponentially. One of the most famous examples of this usage arises when studying the Whitham equations which emerge from the Nonlinear Schr\"odinger equation. One can show in this case that the Benjamin-Feir instability criterion emerges at the point when the characteristics become complex \cite{l65}. This result was the moment when ``the penny dropped'' for Whitham regarding the connection between characteristics and stability \cite{ms16}.

A key development in the setting of this paper is how the dispersionless Whitham system morphs to incorporate dispersive effects at singularities, which was achieved via the work of Bridges et al. \cite{b13,rb16}. The main idea is to modify Whitham's original ansatz with a different set of scalings at points where the Whitham modulation equations develop singularities. In such cases, one instead adopts an ansatz at a new solution of the form
\[
u = \hat{u}\big(\theta+\eps\phi(X,T);k+\eps^2\phi_X(X,T),\omega+\eps^4\phi_T(X,T)\big)+\eps^3 W(\theta,X,T;\eps)\,,
\]
with instead $X = \eps x,\, T = \eps^3 t$. Doing so allows one to resolve the singularity and leads to the emergence of dispersion and the appearance of the well-known nonlinear wave equation, the Korteweg - de Vries (KdV), which governs the evolution of the slowly varying wavenumber. Most strikingly, the KdV which emerges does so in universal form, in the same sense as the Whitham equations. This is to say that its coefficients relate to abstract properties of the original Lagrangian that generates the problem. Extensions to this methodology, for example when resolving additional singularities which may arise, demonstrate how other nonlinear equations such as the two-way Boussinesq and modified KdV (mKdV) emerge. In particular, a recent advancement has shown that when the phase modulation is done in a moving frame, with a speed equal to one of the characteristics emerging from the Whitham system, nonlinear phase dynamics emerge automatically with dispersion \cite{r19}. Further, when the characteristics pass through certain values or coalesce, more complex phase dynamics may occur \cite{br17}. It is for this reason that this paper will adopt the modulation approach in the moving frame, in order to link both the properties of the conservation laws and characteristics of the Whitham equations to the resulting dynamics.

In summary, this paper aims to utilise a modification of the Whitham modulation theory to derive nonlinear dispersive equations to investigate the phase dynamics of the Dysthe system (\ref{DystheA}). This will require a modulation of a two-parameter relative equilibrium, namely the plane wave solution coupled to a uniform current, which adopts a theory adept at treating multiple phases as well as a moving frame. This is achieved by combining existing modulation approaches, and will ultimately be used to show how the KdV equation governing the phase dynamics emerges from the Dysthe equation. Moreover, a secondary aim will be to provide criteria for when one expects the behaviour of defects in the wave to be qualitatively different, occurring precisely when one of these reductions fails to be adequate and should be replaced by another phase dynamical equation. In the setting of this paper, we consider only one example of this, which will focus on how the Gardner, and consequently the mKdV equation, arise whenever the nonlinearity of the KdV equation is sufficiently small and vanishing. This leads to increased nonlinear effects within the evolution of phase defects, which will be highlighted in the body of the paper.

From this phase dynamical description, we aim to provide a possible qualitative picture for the emergence of various coherent structures, such as dark and bright solitary waves. In the study of this paper, we will utilise the various nonlinear dispersive reductions and illustrate how these structures can be interpreted as a deformation under the perturbation of the wave variables by the solitary wave solutions of these systems. This technique has been utilised in other works \cite{boc17,b12,k90,hf14}, however it has yet to be considered in the context of a wave - mean flow system like the Dysthe equation. Thus, another contribution of this work will be to determine how this coupling manifests within this approach as well as to identify its effect on the qualitative predictions. Depending on the phase dynamical description, which itself depends on the properties of the moving frame, we will illustrate that various structures are predicted to form from this viewpoint. Moreover, we will demonstrate that each regime implies that the original wave is affected to differing degrees. Thus, the characteristic speeds from the Whitham equations may in fact be used as a diagnostic regarding the degree in which the phase dynamical description distorts the original wavetrain.

The outline of this paper is as follows. In \S \ref{sec:KdV}, we review how the existing phase modulation approach may be applied to the Dysthe system (\ref{DystheA}) to obtain a KdV equation which governs how the phases of the waves evolve over space and time. This is followed in \S\ref{sec:PD} by an illustration of how such a KdV may be utilised to predict the bifurcation behaviour of the original plane wave solution, with an emphasis on processes that lead to the formation of dark or bright solitary waves. Subsequently, in \S\ref{sec:More}, further phase reductions are discussed, noting when such equations arise and their effect on the plane wave solution. Concluding remarks are presented in \S\ref{sec:CR}.

\section{From Dysthe to KdV}\label{sec:KdV}
There are a multitude of techniques in which to investigate the phase dynamics of equations like (\ref{DystheA}). Primarily, the approach usually taken is to use a standard multiple scales analysis and solve at each order of the small parameter \cite{k90,b12}. A variant of this, and the closest reduction procedure to the modulation considered here, is to undertake this approach after transforming the system using the Madelung transform \cite{boc17,fs02,hf14}. This approach is utilised in appendix \ref{App:PD} to derive the KdV and show that it agrees with that obtained via the modulation methodology. Prior to this paper, it does not appear that such a study using the Madelung approach has been undertaken in the context of wave - mean flow systems such as the Dysthe equation, and so the reduction using this method is also another novelty of this paper. Within this paper the focus will be on the modulation approach, with the details of how the KdV equation may be obtained using this procedure appearing in appendix \ref{App:Mod}. This method allows one to draw a connection between not only the conservation laws the Dysthe system possesses, but also highlights how the hyperbolicity of the system plays a role in the evolution of the phase defects.

In order to apply the modulation approach detailed in appendix \ref{App:Mod} to the Dysthe system (\ref{DystheA}), we must first note that it is generated by a Lagrangian, namely
\begin{multline}\label{Lagrangian}
\mathcal{L} = \iint\Bigg[\frac{i}{2}(AA^*_t-A^*A_t)+|A_x|^2+\frac{i \beta}{4}|A|^2(AA^*_x-A^*A_x)\\
-\alpha |A|^2\left.\Phi_x\right\vert_{z=0}-\frac{1}{2}\int_{-h}^0\Phi_x^2+\Phi_z^2\, d z\Bigg]\ dx\, dt\,.
\end{multline}
In order to discuss how defects in the phase evolve for uniform plane waves with a current in this equation, we consider the relative equilibria associated with the product of an affine and toral symmetry. The first of these arises from the fact that one may add arbitrary constants to the velocity potential $\Phi$ and leave the Lagrangian invariant, and the second is the one associated with the $S^1$ symmetry of the amplitude $A$. This corresponds to the uniform wavetrain state in $A$ and the uniform flow solution of $\Phi$, which explicitly is given by
\begin{equation}\label{Plane-Sol}
A = A_0 e^{i \theta}\,, \quad \Phi = u_0x+\gamma t\,, \quad \mbox{with} \quad |A_0|^2 =\Psi =  \frac{\alpha u_0-k^2-\omega}{1+\beta k}\,.
\end{equation}
The amplitude $|A_0|^2$ must be positive and nonsingular, meaning that $k \neq -\frac{1}{\beta}$.

For the analysis of appendix \ref{App:Mod} we require the relevant conservation law vectors. These arise from averaging the Lagrangian (\ref{Lagrangian}) over a period of the wave, as determined in (\ref{cons-laws-app}) and can be readily determined as
\[
{\bf A}
 = \begin{pmatrix}
 \widehat{\mathcal{L}}_\omega\\
  \widehat{\mathcal{L}}_{\gamma}
 \end{pmatrix}
 = \begin{pmatrix}
\Psi\\
0
\end{pmatrix}\,, \quad 
{\bf B}
 = \begin{pmatrix}
 \widehat{\mathcal{L}}_k\\
  \widehat{\mathcal{L}}_{u_0}
 \end{pmatrix}
 = \begin{pmatrix}
2k\Psi+\frac{\beta}{2}\Psi^2\\
-hu_0-\alpha \Psi
\end{pmatrix}\,.
\]
The first row of these vectors are associated with the conservation of wave action of the periodic wave and is akin to the conservation of mass for the wave itself. The second emerges as the conservation of mass of the flow beneath the wave, albeit with a negative sign. This is due to the way one obtains it via the averaging principle as detailed in (\ref{cons-laws-app}).
With these quantities in hand, we are in a position to compute both the criterion for emergence and the coefficients of the KdV associated with this wavetrain. The relevant criticality condition for this to emerge in a moving frame, with characteristic speed $c$, from the solution (\ref{Plane-Sol}) is given by (\ref{Del-c-app}) in appendix \ref{App:Mod}. For the Dysthe equation, this requires that the determinant of the Jacobian
\[
{\bf E}(c) = \frac{1}{1+\beta k}
\begin{pmatrix}
2(1+\beta k)\Psi-(2k+c+\beta \Psi)^2&\alpha(2k+\beta \Psi+c)\\
\alpha(2k+c+\beta \Psi)&-h(1+\beta k)-\alpha^2
\end{pmatrix}\,,
\]
vanishes. This is the case whenever the characteristic speed satisfies
\begin{equation}\label{c-cond}
\Delta(c)=hc^2+2h(\beta\Psi+2k)c+\beta^2h\Psi-2\big(h(1-\beta k)+\alpha^2\big)\Psi+4hk^2 = 0\,,
\end{equation}
and so the characteristic speeds are given by
\begin{equation}\label{c-defn}
c = -2k-\beta \Psi \pm \sqrt{2\Psi\bigg(\beta k+1+\frac{\alpha^2}{h}\bigg)}\,.
\end{equation}
The necessity for hyperbolicity requires that $k>-\frac{1}{\beta}-\frac{\alpha^2}{\beta h}$.
In such cases where (\ref{c-cond}) is satisfied and $c$ is real, we can define the relevant eigenvector of ${\bf E}(c)$ required for the theory, $\be$, as
\[
\be = 
\begin{pmatrix}
\zeta_1\\
\zeta_2
\end{pmatrix} 
= \frac{1}{1+\beta k} 
\begin{pmatrix}
h(1+\beta k)+\alpha^2\\
\alpha(2k+c+\beta \Psi)
\end{pmatrix}\,,
\]
We may now compute the coefficients of the resulting KdV equation, which is done by following the methodology in \cite{br18} and appendix \ref{App:Mod}. The first to be computed is the coefficient of the time derivative term, giving that
\[
\be^T{\bf E}'(c)\be = -\frac{2\big((\beta  k+1)h+\alpha^2\big)(\beta\Psi+c+2k)h}{(bk+1)^2}\,.
\]
The next calculation is for the coefficient of the nonlinear term, and gives that
\[
\begin{split}
\be^T{\bf H}(\be,\be)=\frac{3\big((\beta  k+1)h+\alpha^2\big)(\beta\Psi+c+2k)h\big(h(\beta^2\Psi+\beta c-2)-2\alpha^2 \big)}{(bk+1)^3}\,.
\end{split}
\]
The final coefficient required is that of the dispersive term. Typically within the modulation approach this is acquired by undertaking a Jordan chain analysis, however it is in fact more readily obtained from the dispersion relation computed about the solution (\ref{Plane-Sol}), as is typically done in hydrodynamic settings \cite{a94,g77,gos98}. This is since the dispersion relation for the KdV must match up with the dispersion relation from the original problem (\ref{DystheA}) in the long wave limit. This can be found from a simple linear analysis of (\ref{DystheA}) about the plane wave solution (\ref{Plane-Sol}), and leads to the dispersion relation
\[
\sigma = -(2k+\beta \Psi+c)\kappa\pm \kappa\sqrt{2 \Psi(1+\beta k)+\kappa^2+\frac{\alpha^2\Psi}{h}\kappa\coth(\kappa h)}\,.
\]
Thus, the relevant coefficient of dispersion arises from the cubic term in $\kappa$ in the long wave expansion of the above relation, multiplied by the coefficient of the KdV's time derivative term. This leads to the required coefficient of dispersion
\[
\frac{1}{6}\left.\sigma_{\kappa\kappa\kappa}\right\vert_{\kappa=0}\be^T{\bf E}'(c)\be =-\frac{\big((\beta  k+1)h+\alpha^2\big)h(3+2\alpha^2h\Psi)}{3(bk+1)^2}\,.
\]
This can be verified by following the approach outlined in appendix \ref{App:PD}, up to a scaling factor.
Combining the above results and simplifying gives the relevant KdV as
\begin{multline}\label{KdV}
(2k+\beta\Psi+c)\bigg(U_T- \frac{3}{2(1+\beta k)}\big(h(\beta^2\Psi+\beta c-2)-2\alpha^2 \big)UU_X\bigg)\\
+\frac{1}{6}(3+2\alpha^2h\Psi)U_{XXX} = 0\,.
\end{multline}
One can notice that the mean flow effects, characterised by $\alpha$, appear explicitly within the dispersive coefficient, whereas both the characteristic speed $c$ and the self steepening effects determined by $\beta$ appear within the coefficients of the time and nonlinear terms. Thus, these effects have a non-negligible effect on the phase dynamics which emerge from the plane wave solution. The aim now will be to investigate how these effects influence the phase dynamics and lead to the emergence of coherent, localised structures from the original plane wave.

\subsection{The Evolution of Phase Defects}\label{sec:PD}
With the relevant KdV (\ref{KdV}) in hand, we may now discuss how it may be used to determine the bifurcating behaviour of the periodic wave solution (\ref{Plane-Sol}). There are a large family of solutions admitted by the KdV equation, such as cnoidal waves \cite{d77} and multipulse solutions \cite{h71,ms93}, which could be utilised to explore the phase dynamics as predicted by (\ref{KdV}).  However, for simplicity, we will focus on how the KdV equation in this context can describe the formation of dark and bright solitary waves from the uniform plane wave solution. The strategy to illustrate this is to use the solitary wave solution to the KdV equation (\ref{KdV}),
\[
U(\xi) = a\,{\rm sech}^2\bigg(\frac{\xi}{W}\bigg)\,,
\]

with $\xi = X-VT$,and where the amplitude $a$ and width $W$ are given by
\begin{equation}\label{amp-width-phi}
\begin{split}
a =& -\frac{2V(1+\beta k)}{\big(h(\beta^2\Psi+\beta c-2)-2\alpha^2 \big)}\,, \qquad W = 2 \ \sqrt{\frac{3+2\alpha^2h\Psi}{6(2k+\beta\Psi+c) V}} \,.
\end{split}
\end{equation}
 This is then used to reconstruct the solution to the Dysthe equation according to the ansatz
\begin{equation}\label{Reconst}
\begin{split}
A &= A_0(k+\eps^2\zeta_1U,u_0+\eps^2\zeta_2U,\omega+c\eps^2\zeta_1U-V\eps^4\zeta_1U)e^{i(\theta+\eps \zeta_1\phi)}\,,\\
& \hspace{1cm}\mbox{where} \quad  \phi = \int U d\xi\,, =aW \tanh\bigg(\frac{\xi}{W}\bigg)
\end{split} 
\end{equation}
which is utilized to derive the KdV (\ref{KdV}). We then determine how the original solution (\ref{Plane-Sol}) is impacted. This, to leading order, has the corresponding effect on the envelope:
\begin{equation}\label{amp-pert}
\begin{split}
|A|^2 &= \Psi+\eps^2 \ \frac{2(c+2k+\beta\Psi)V}{\beta^2 \Psi+\beta c-2-\frac{2 \alpha^2}{h}}{\rm sech}^2 \bigg(\frac{\xi}{W}\bigg)+\mathcal{O}(\eps^4)\,.
\end{split}
\end{equation}
It is clear at this stage that both the mean flow and steepening influence the amplitude and width of the solitary wave and will lead to a wide range of possible localised structures. However, we note that the current $u_0$ does not explicitly enter any of these expressions, and so for a fixed amplitude $\Psi$ the resulting dynamics is independent of the mean flow velocity, although the strength of these effects does influence the observed phenomenon.

We can already determine a great deal analytically. In particular, we can see that as the strength of the mean flow $\alpha$ is increased the width of the solution $U$, and thus the width of the observed disturbance to the wave,  increases. This is the case for either choice of the characteristic speeds, as made clear by (\ref{c-defn}) which highlights that
\[
c   +2k+\beta \Psi =  \pm \sqrt{2\Psi\bigg(\beta k+1+\frac{\alpha^2}{h}\bigg)}\,,
\]
and so this only affects the choice of sign for $V$.
Thus the mean flow is the dominant effect in the width, as the steepening only asymptotically decreases the width as $\beta^{-\frac{1}{2}}$ whereas the mean flow causes this to grow as $\alpha$.

The effect of the phase dynamics on the amplitude of the plane wave is less clear and involves an interplay between steepening and depth. However, by studying (\ref{amp-pert}), one is able to see that overall the effects of steepening and the mean flow decrease the resulting amplitude of the structure which forms, and the asymptotic decay is algebraic of the order $\alpha^{-1},\,\beta^{-1}$. As such, one expects the largest amplitudes for these disturbances to occur for $\alpha = 0$ and for values of $\beta$ approaching the singularity in the amplitude of the solution $U$ as given in (\ref{amp-width-phi}). However in this proximity the KdV equation (\ref{KdV}) begins to become an invalid model of the phase dynamics and other descriptions become operational, as will be described within the next section, so the discussion here will not apply to such choices of $\beta$ that are sufficiently close to these points.

In summary, this approximate picture provides us with a qualitative description for how these localised structures appear and what form these are expected to take. In the presence of strong mean flow interactions between the wave and the current, the distortion to the envelope is expected to have a longer range. When the wave is subjected  to increased self-steepening effects, the magnitude of the disturbance is muted so long as a degeneracy in the derived KdV equation is not approached in parameter space. Therefore, the KdV dynamics derived in this setting suggest that the original plane wave is in fact stable to the majority of defects which can form, and their effect on the wavetrain is very limited.

\begin{figure}[!ht]
\centering
\begin{subfigure}{0.49\textwidth}
\includegraphics[width=\textwidth]{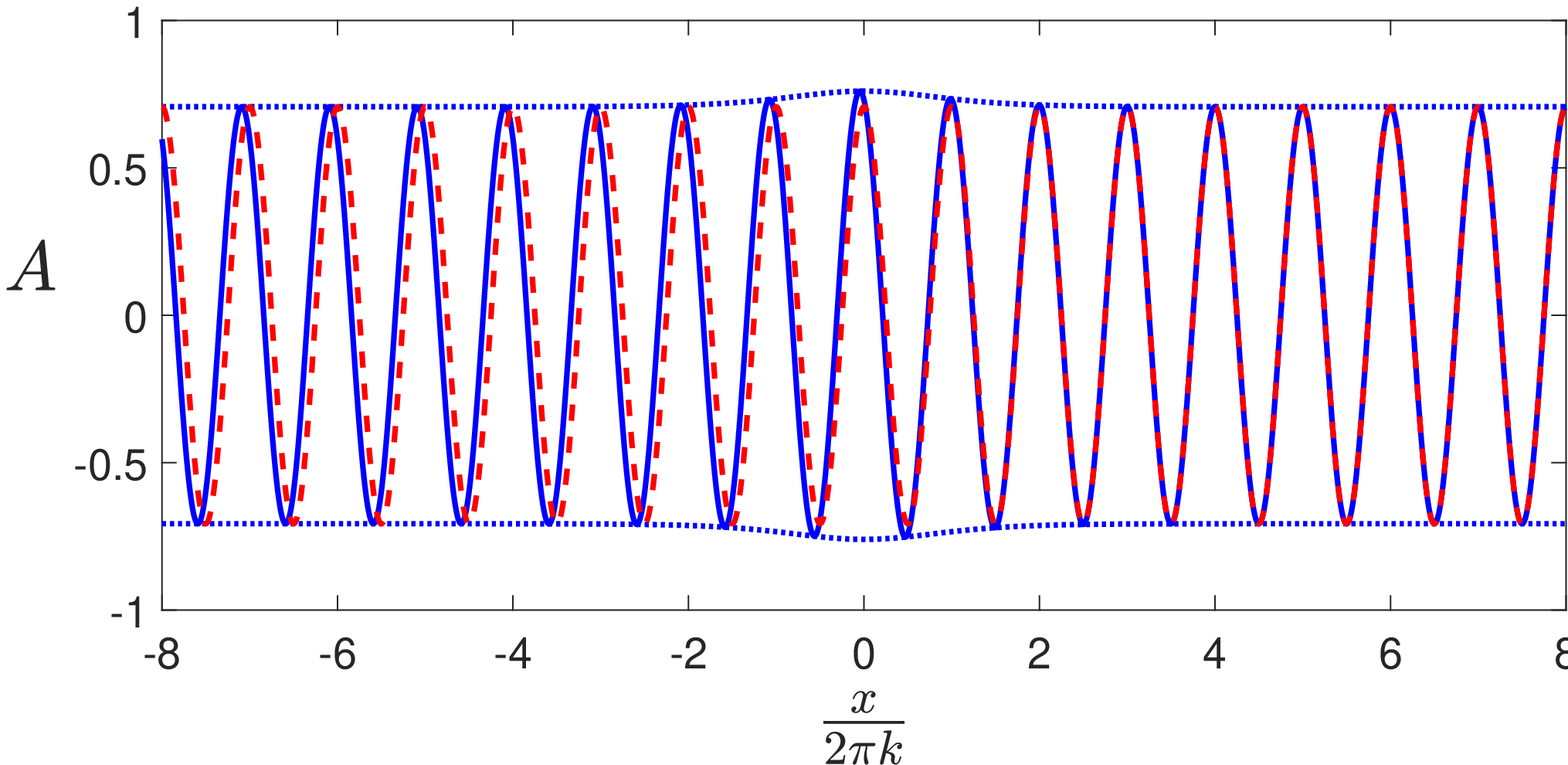}
\caption{}
\end{subfigure}
\begin{subfigure}{0.49\textwidth}
\includegraphics[width=\textwidth]{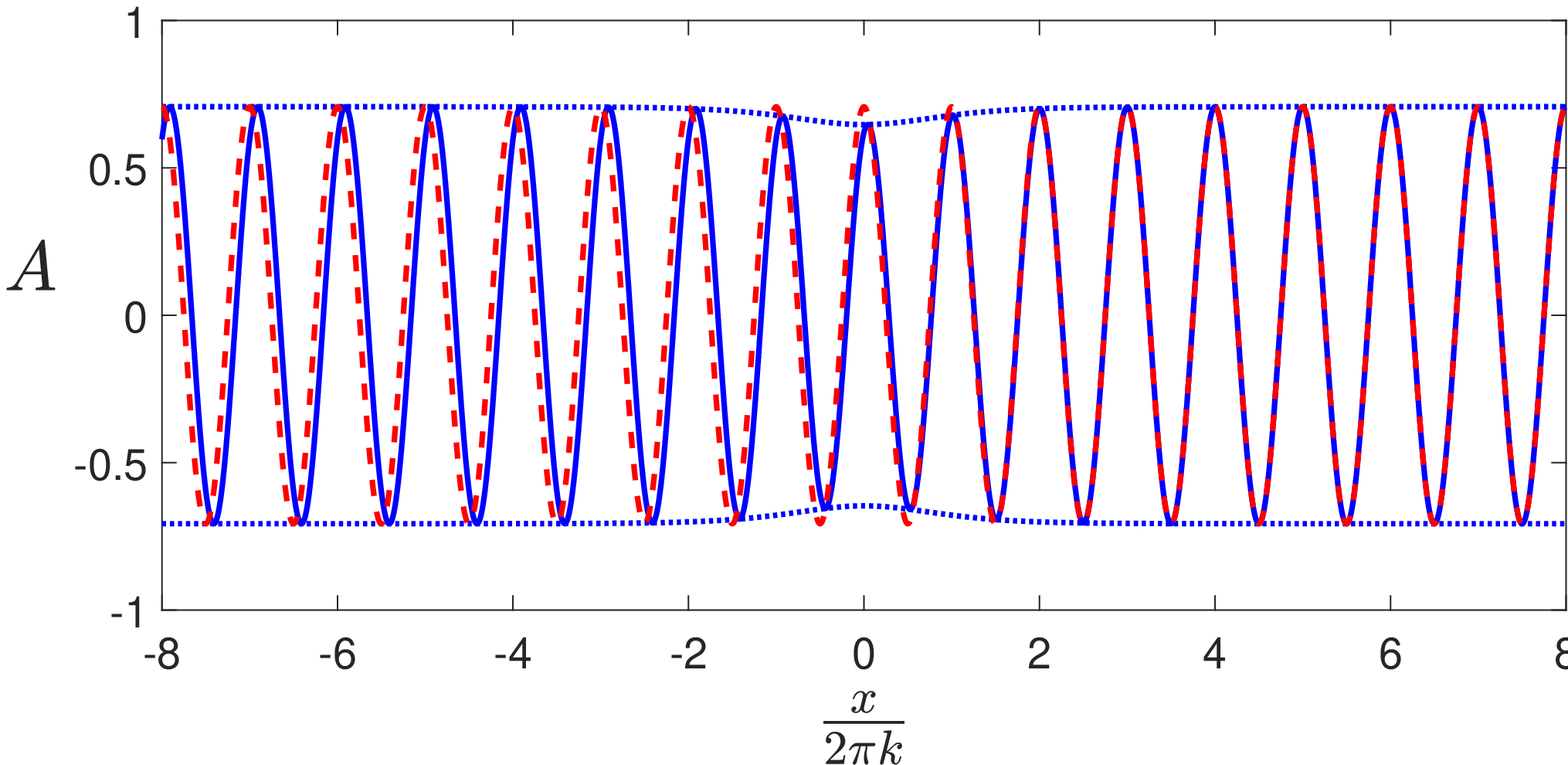}
\caption{}
\end{subfigure}
\begin{subfigure}{0.49\textwidth}
\includegraphics[width=\textwidth]{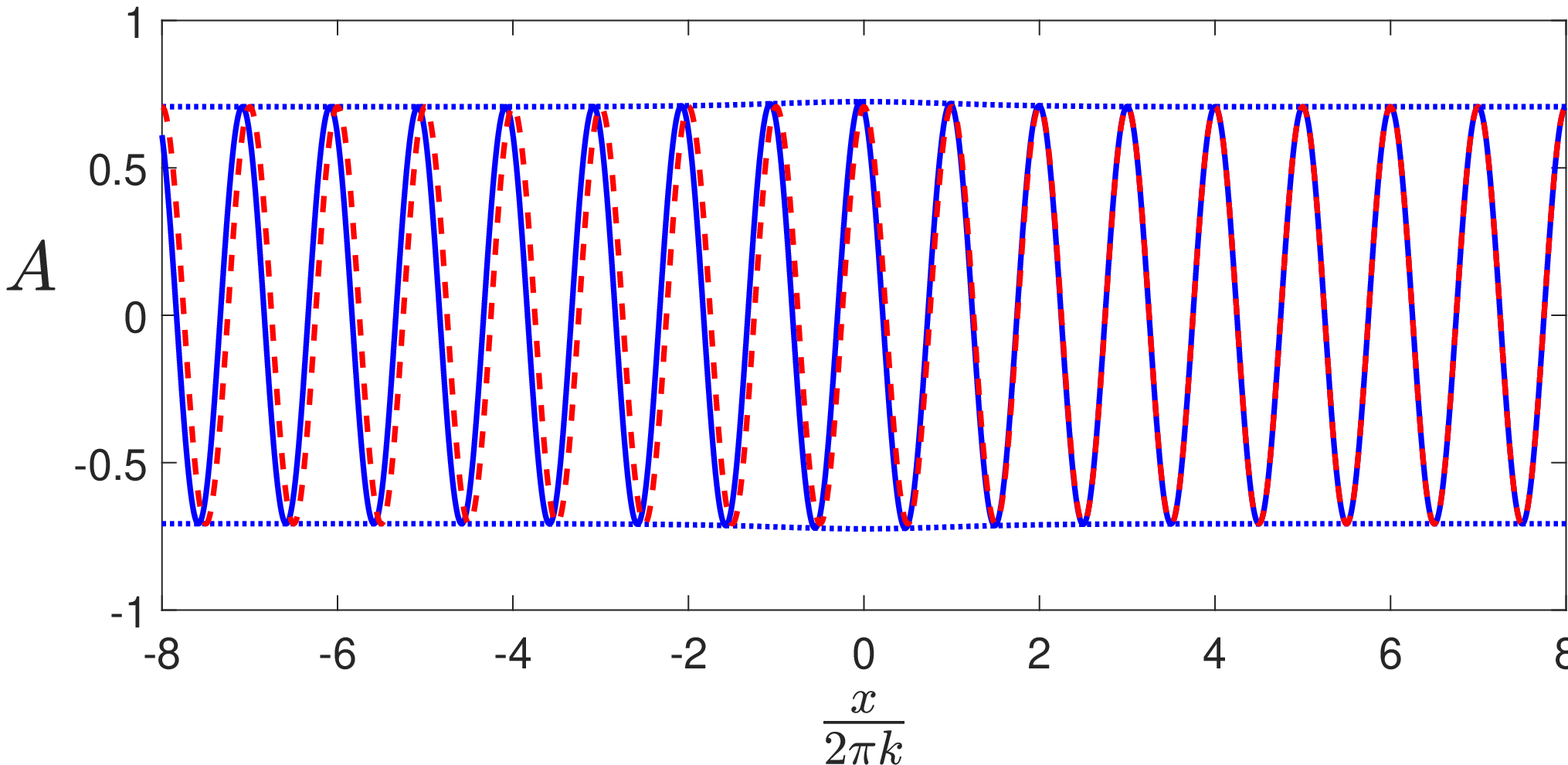}
\caption{}
\end{subfigure}
\begin{subfigure}{0.49\textwidth}
\includegraphics[width=\textwidth]{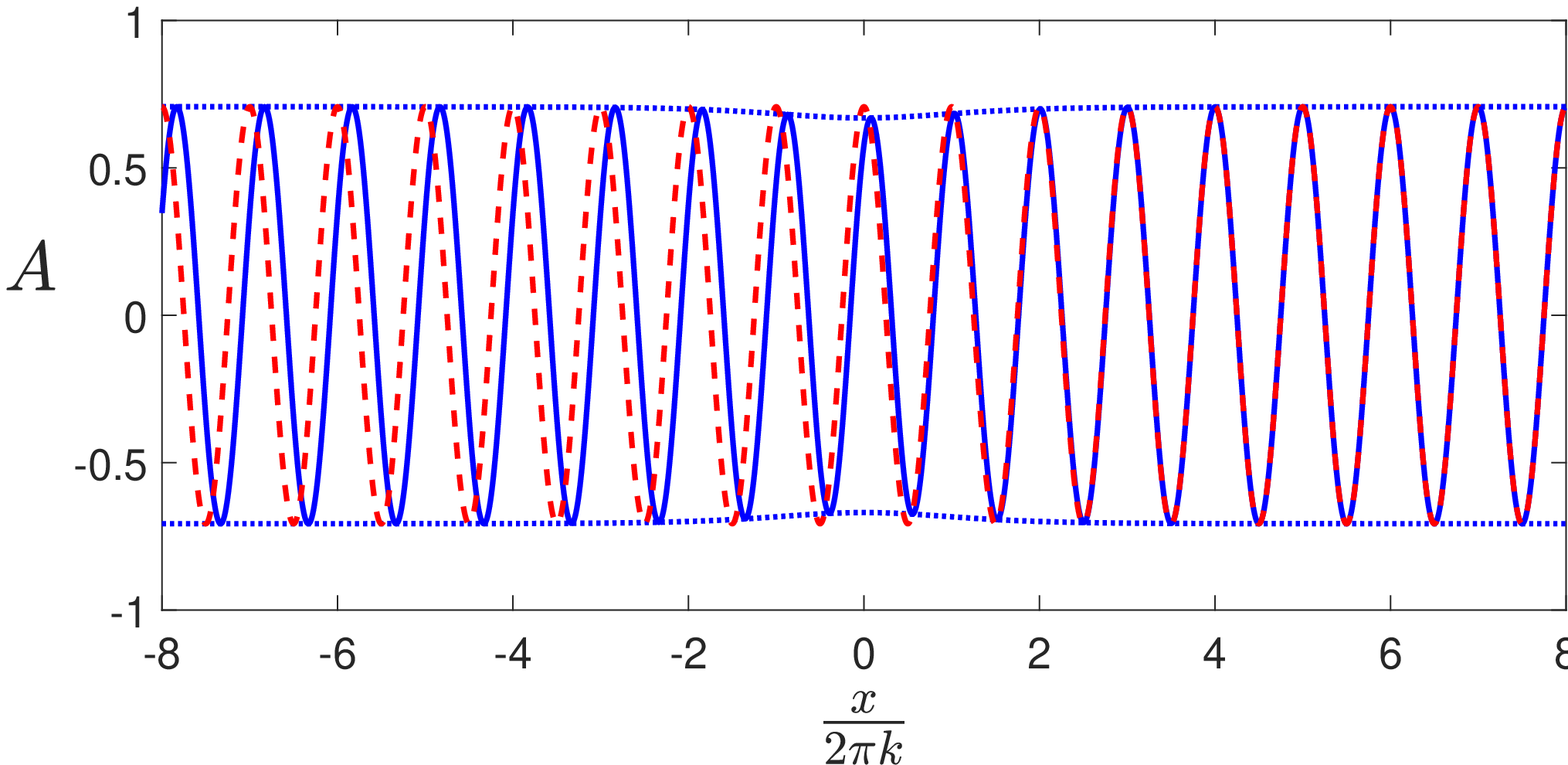}
\caption{}
\end{subfigure}
\begin{subfigure}{0.49\textwidth}
\includegraphics[width=\textwidth]{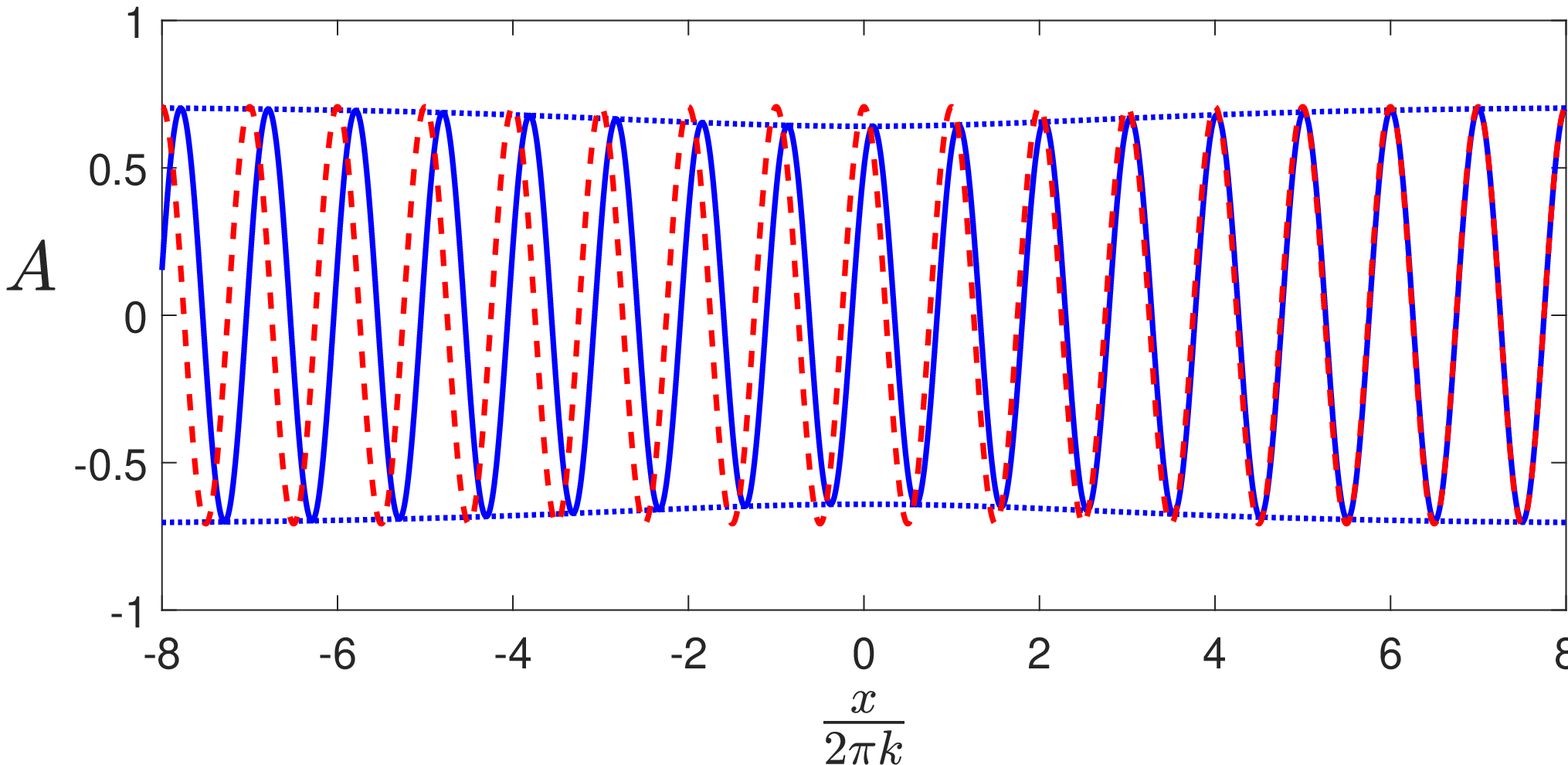}
\caption{}
\end{subfigure}
\begin{subfigure}{0.49\textwidth}
\includegraphics[width=\textwidth]{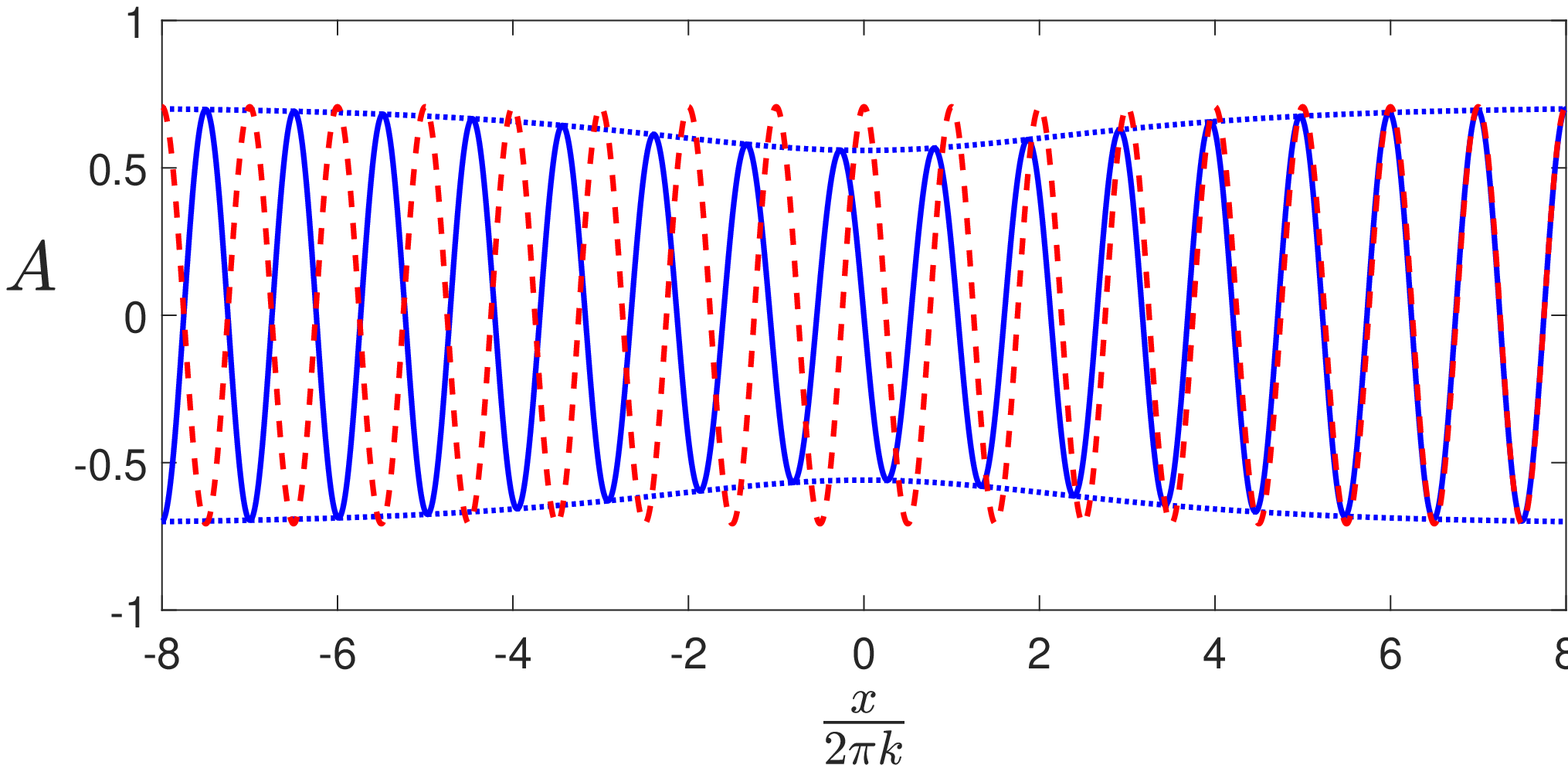}
\caption{}
\end{subfigure}
\begin{subfigure}{0.49\textwidth}
\includegraphics[width=\textwidth]{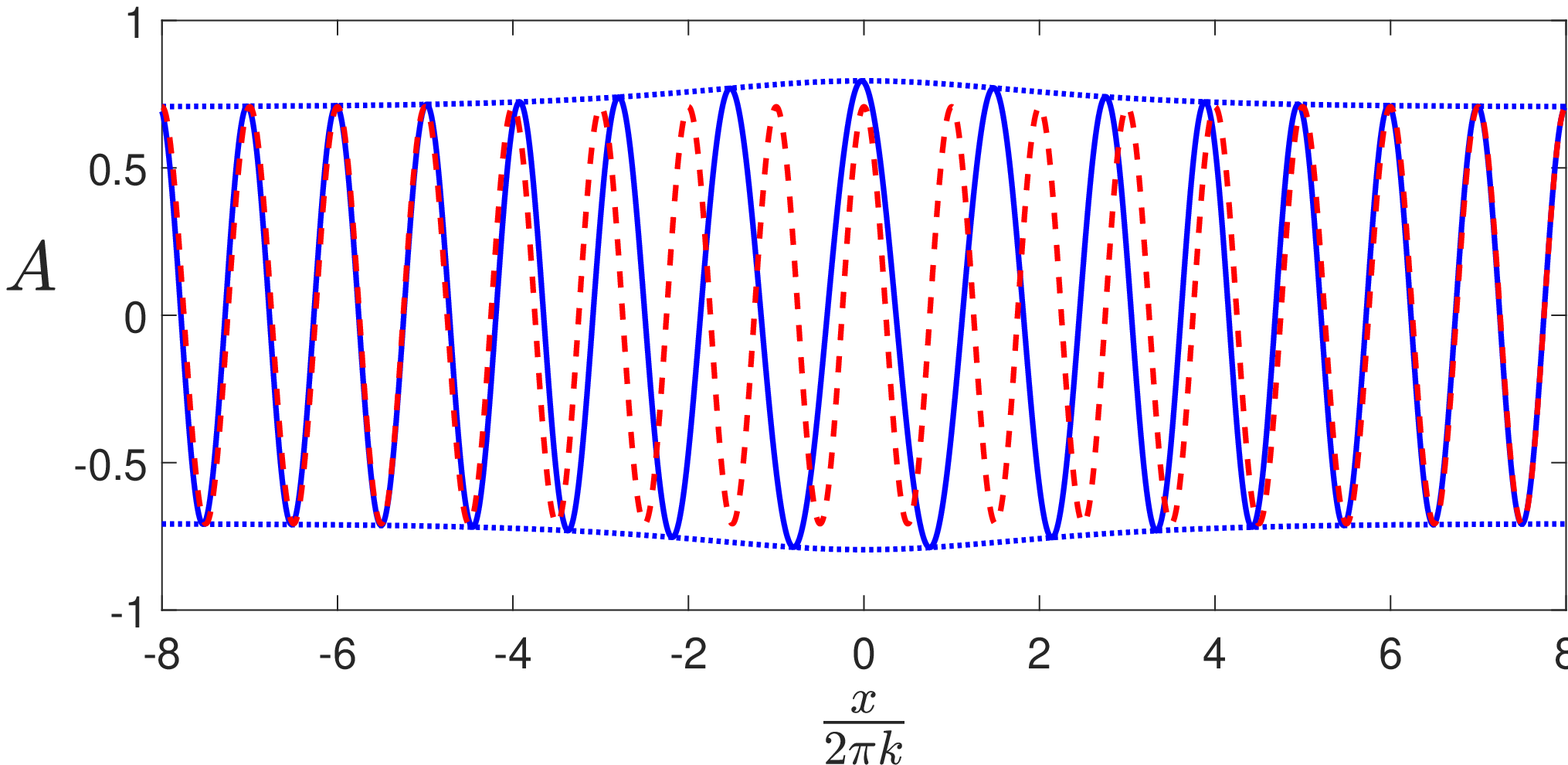}
\caption{}
\end{subfigure}
\begin{subfigure}{0.49\textwidth}
\includegraphics[width=\textwidth]{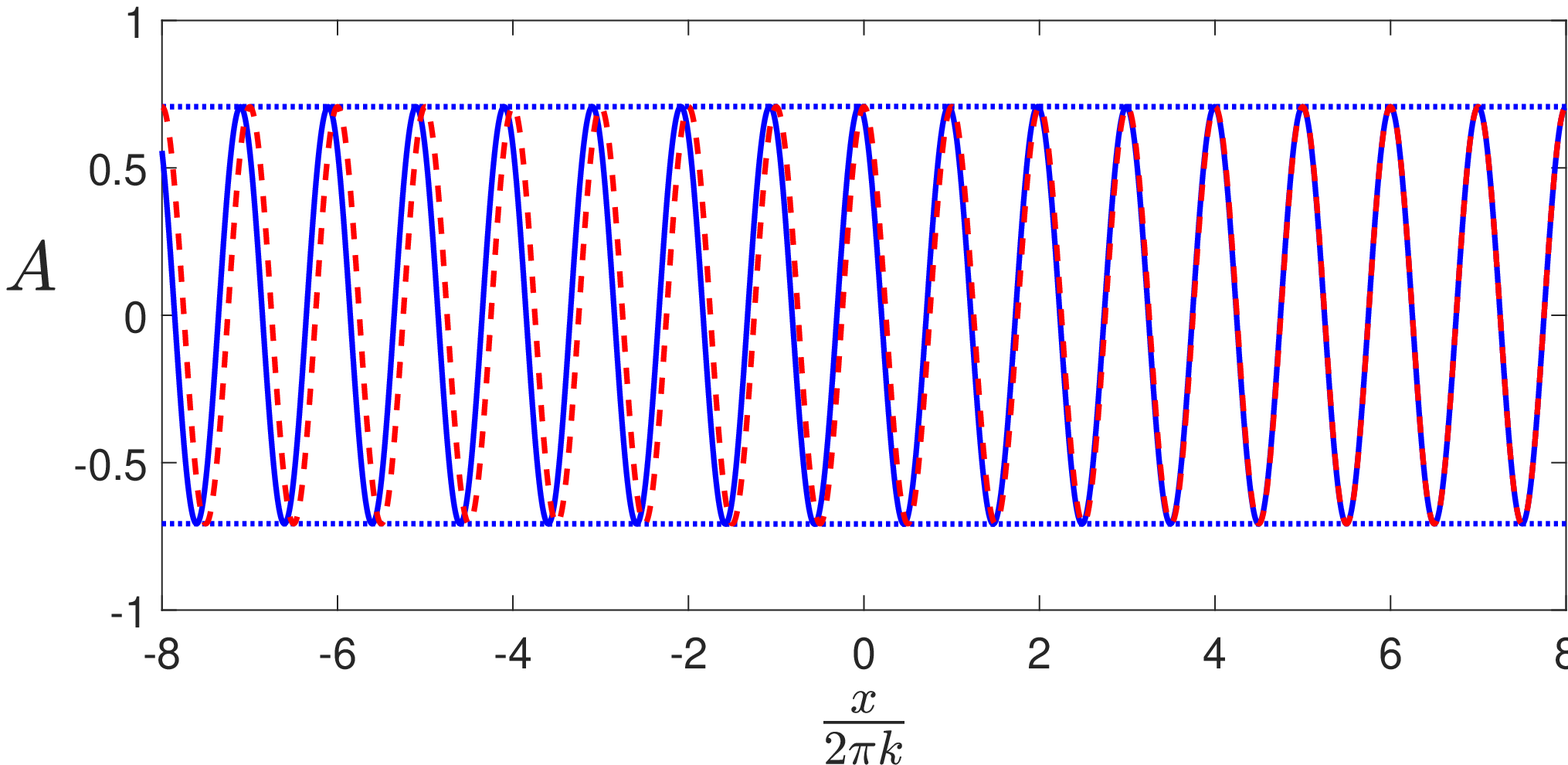}
\caption{}
\end{subfigure}
\caption{A comparison between the resulting bifurcation of the plane wave solution (\ref{Plane-Sol}) under the phase dynamics for the fast (left column) and slow (right column) characteristic speeds for $\eps = 0.2$, $k = 1,\,\omega = -\frac{3}{2},\,h = 4,\, V = 1$ and (a)-(b) $\alpha = \beta = 0$, (c)-(d) $\alpha = \beta = 1$, (e)-(f) $\alpha = -4,\,\beta = -\frac{3}{2}$ and (g)-(h) $\alpha = 4,\,\beta = 8$. The dashed red lines indicate the original plane wave, whereas the solid blue lines denote the solution reconstructed via (\ref{Reconst}), with the dotted blue lines corresponding to the envelope for this solution.}
\label{fig:comp}
\end{figure}

We combine these findings to illustrate what effects the mean flow and self-steepening have upon the bifurcation behaviour of the original plane waves. This is done by setting $\alpha = \beta = 0$ and in essence becomes a study of the phase dynamics of the Nonlinear Schr\"odinger equation \cite{k90,b12}. This is then compared to cases where $\alpha,\,\beta$ are nonzero, but with the same choice of wavenumber and frequency so that the effects of the mean flow and steepening can be assessed. The mean flow $u_0$ is chosen so that the amplitudes in all cases are the same. An example of such a comparison is depicted in figure \ref{fig:comp}. 

We now discuss the results of this investigation. One can observe that the mean flow and steepening effect can lead to qualitative differences in the dynamics, changing the polarity of the resulting structure from dark to bright, and vice-versa. This is to be expected, since the steepening effects enter the nonlinear coefficient and allow for it to change sign. Moreover, we can see that these effects can also lead to the previously discussed suppression of the solitary structures, suggesting that the original wave persists even in the presence of the nonlinear phase dynamics and hints at increased stability in these regimes. Additionally, it becomes clear that the presence of mean flow and self-steepening within the system can lead to increased nonlinear effects in the predicted bifurcating behaviour, where these disturbances are enhanced. Overall, the presence of both of these effects leads to a greater range of behaviours for the localised structures.

\section{Further Singularities and Higher Order Phase Dynamics}\label{sec:More}
Although the KdV itself provides some insight into the evolution of defects and the formation of coherent structures from the uniform wavetrain, there are parameter values for which the KdV stops being operational due to vanishing terms. At such points, there is the potential for even more interesting nonlinear phenomen\red{a} to emerge in a way analogous to a secondary instability via higher order phase equations. We discuss one such case of this below, leading to the modified KdV equation, and by extension the Gardner equation, with a discussion of their effects on the evolution of the original plane wave solution.

\subsection{Modified KdV}
The singularity of interest is the case where only the quadratic nonlinearity vanishes, which is whenever the characteristic speed takes the value
\begin{equation}\label{c-mkdv}
c = \frac{1}{\beta}\bigg(2-\beta^2\Psi+\frac{2\alpha^2}{h}\bigg)\,.
\end{equation}
For this to occur, the amplitude must satisfy
\begin{equation}\label{mKdV-cond}
\Psi = \frac{2}{\beta^2}\bigg(\beta k+1+\frac{\alpha^2}{h}\bigg)\,.
\end{equation}
Using this value of the amplitude in (\ref{c-mkdv}) gives the much simpler condition that $c = -2k$. From this, we note that without the moving frame one would have to impose $k=0$ to obtain the mKdV from the Dysthe equation, limiting its applicability as well as its ability to emerge from plane wave solutions.

In such cases where (\ref{c-mkdv}) holds, a rescaling of the modulation approach must occur in order to reintroduce a nonlinear term into the analysis. The subsequent modulation reduction procedure then follows very similarly to \cite{r18}, with the modifications for the moving frame appearing in other works \cite{br17,br18} and appendix \ref{App:Mod}.
Alternatively, one considers a rescaling of the ansatz used in appendix \ref{App:PD} as is done to obtain the mKdV in similar contexts (for example, \cite{s73,ky78}).
By following either procedure, one obtains the modified KdV equation
\begin{equation}\label{mKdV}
\beta\Psi U_T+\frac{3\beta^2\Psi^2}{2(1+\beta k)^2}U^2U_X+\frac{3+2\alpha^2h\Psi}{6}U_{XXX} = 0\,.
\end{equation}
Of note is that the effects of steepening now have an increased role in the dynamics through the coefficient of the cubic nonlinearity. 

In order to investigate dark and bright solitary waves in this regime, the relevant solitary wave solution family to the above is given by
\begin{equation}\label{sol-mKdV}
U = \pm\sqrt{\frac{V(1+\beta k)^2}{\beta\Psi}} \ {\rm sech}\bigg(\sqrt{\frac{6  \beta V \Psi}{3+2\alpha^2\Psi h}} \ \xi\bigg)\,.
\end{equation}
The requirement that this solution be real imposes that $\beta V>0$.
Note that solutions of both polarities are permissible from the phase dynamics now, and so one expects both dark and bright solitary waves to emerge in this scenario at the same parameter values. The solution is then reconstructed according to
\begin{equation}\label{ansatz-mKdV}
A = A_0(k+\eps\zeta_1U,u_0+\eps\zeta_2U,\omega+c\eps\zeta_1U-V\eps^3\zeta_1U)e^{i(\theta+\zeta_1\phi)}\,,
\end{equation}
where the relevant scalings in $\eps$ are chosen so that the corresponding modulation approach leads to the mKdV \cite{r18a}. To leading order, the effects on the envelope of the wave can be deduced via a Taylor expansion, revealing that
\[
|A|^2 = \Psi\pm\eps \sqrt{\frac{2h(\beta h k+h+\alpha^2)V}{\beta}}{\rm sech}\bigg(\sqrt{\frac{6  \beta V \Psi}{3+2\alpha^2\Psi h}} \ \xi \bigg)+\mathcal{O}(\eps^2)\,.
\]
From this, we can make some inferences regarding how the mean flow and steepening alter the phase dynamical observations. Firstly, the fact that the solution admitted is sech to a unitary power reveals that the dark or bright structures are already expected to be wider than those observed under the KdV dynamics. We can also see that $\beta$ now has a decreased role in the amplitude of the disturbance, and instead the mean flow and depth become the main contributing factors to its size. Moreover, the expansion itself leads to a correction an order lower than that of the KdV case, and so the effects of this are expected to be more readily observed. The width of the localised structure is very similar to that discussed in the KdV equation, which is expected as the linear dispersive properties of the phase dynamics have not changed, and thus has the same asymptotic decay properties.

\begin{figure}[!ht]
\centering
\begin{subfigure}{0.49\textwidth}
\includegraphics[width=\textwidth]{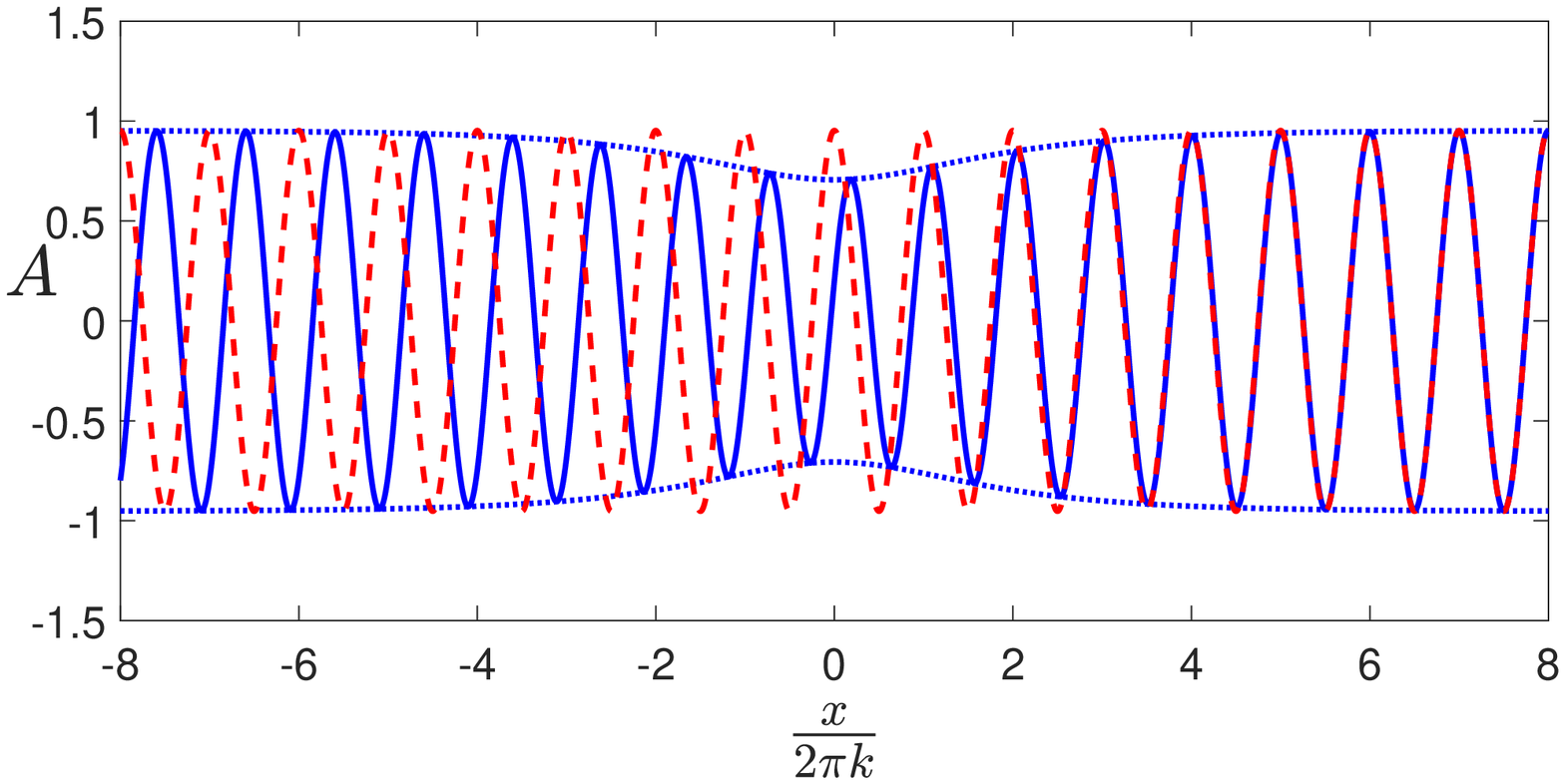}
\caption{}
\end{subfigure}
\begin{subfigure}{0.49\textwidth}
\includegraphics[width=\textwidth]{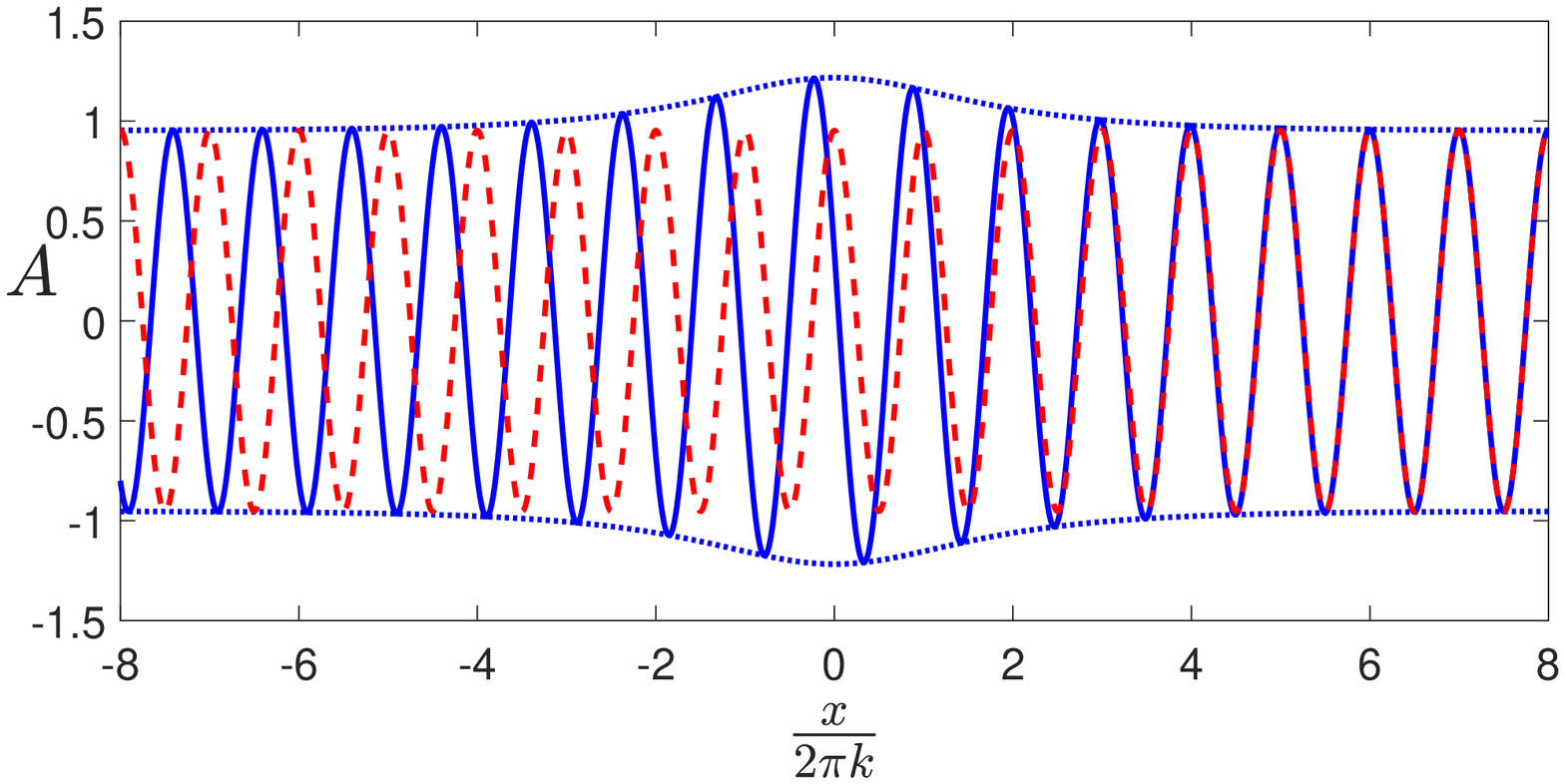}
\caption{}
\end{subfigure}
\begin{subfigure}{0.49\textwidth}
\includegraphics[width=\textwidth]{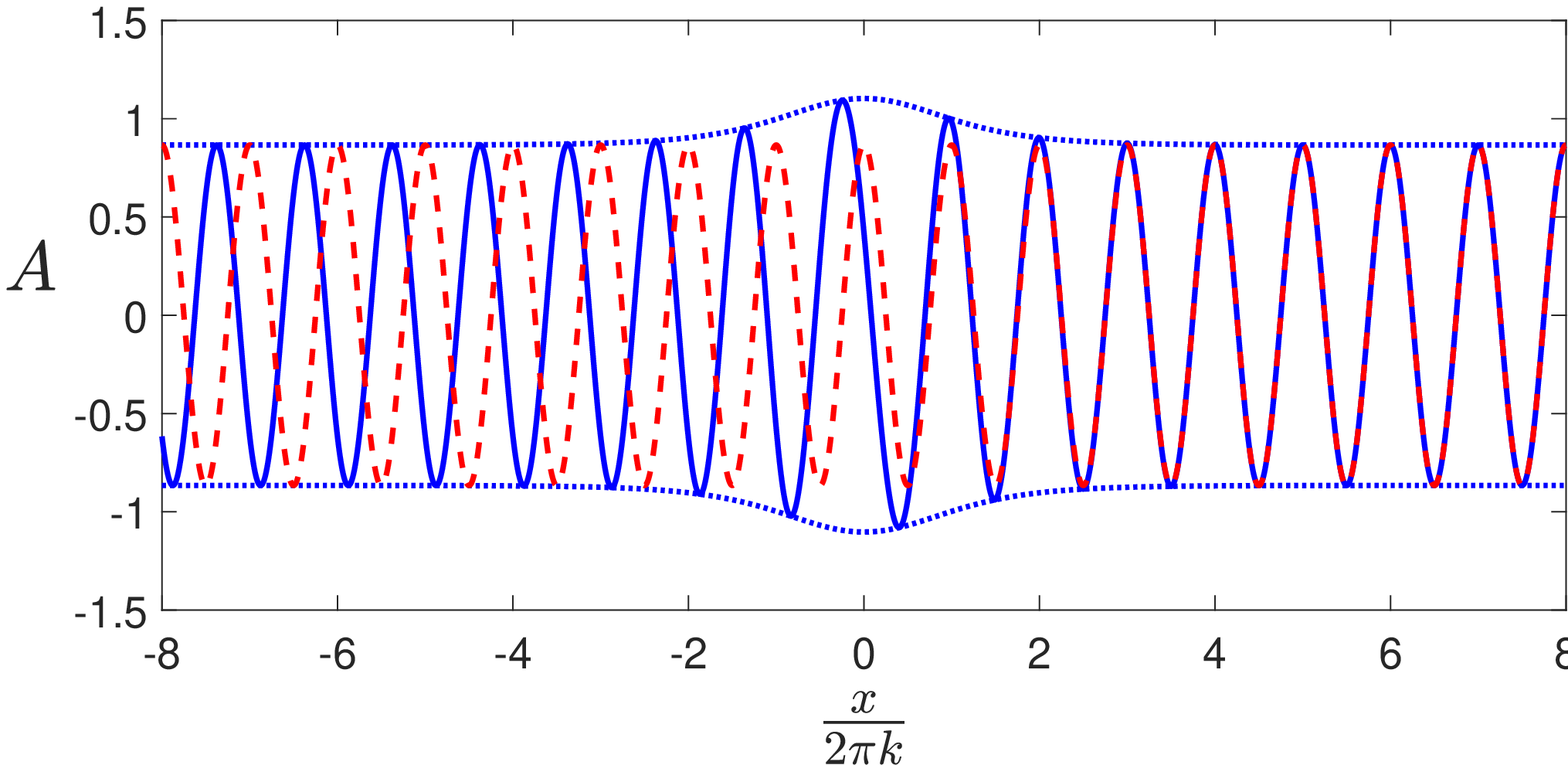}
\caption{}
\end{subfigure}
\begin{subfigure}{0.49\textwidth}
\includegraphics[width=\textwidth]{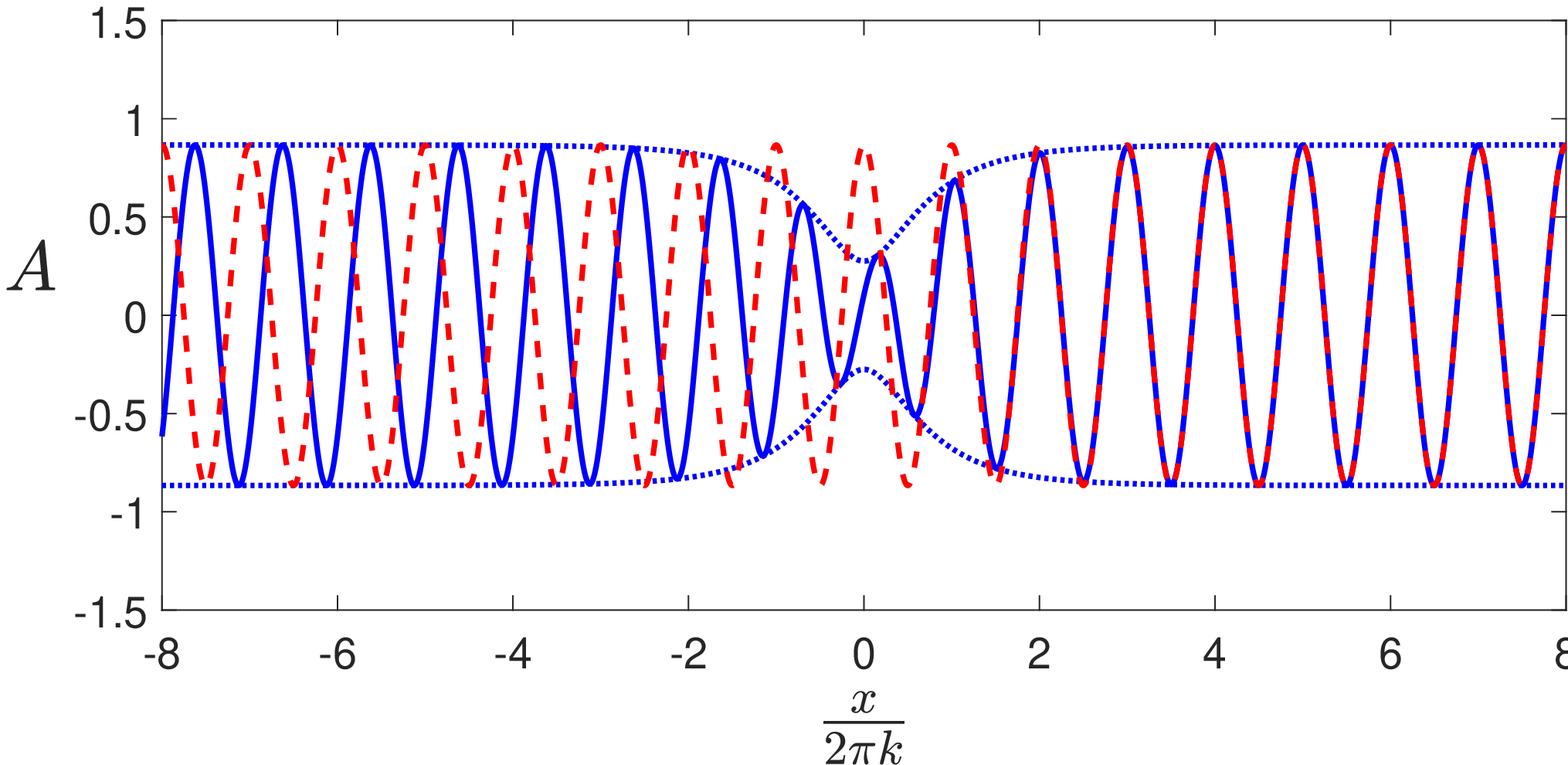}
\caption{}
\end{subfigure}
\caption{Examples of how the solution (\ref{sol-mKdV}) distorts the original plane wave for (a)-(b) $\alpha = -0.5,\,\beta = 3,\,k=1,\,\omega = -4.32,\,h = 3,\,u_0 = 1,\,c=1,\, \eps = 0.06$ and (c)-(d) $\alpha = 0.7,\,\beta = -1.2,\,k=1.2,\,\omega = -0.69,\,h = 0.5,\,u_0 = 1,\,c=-1.6,\,\eps = 0.2$ for both positive (left) and negative (right) polarities.}
\label{fig:mkdv}
\end{figure}

These observations are confirmed by the examples of how these solutions modify the original plane wave\red{,} depicted in figure \ref{fig:mkdv}. Overall, the reconstructions in this case tend to have a more pronounced but typically broader effect on the plane wave than the KdV equation, causing more apparent versions of the emergent dark and bright solitary waves. This is expected, as the mKdV soliton is both narrower and larger in amplitude than the one which is admitted by the KdV. Moreover, the scalings within the ansatz to obtain the mKdV are larger than the KdV case. Thus, the resulting bifurcation that the phase dynamics predicts should be of higher magnitude, as well as being sharper in regimes where the mKdV (\ref{mKdV}) is operational.

\subsection{Gardner Equation}

\begin{figure}[!ht]
\centering
\begin{subfigure}{0.7\textwidth}
\includegraphics[width=\textwidth]{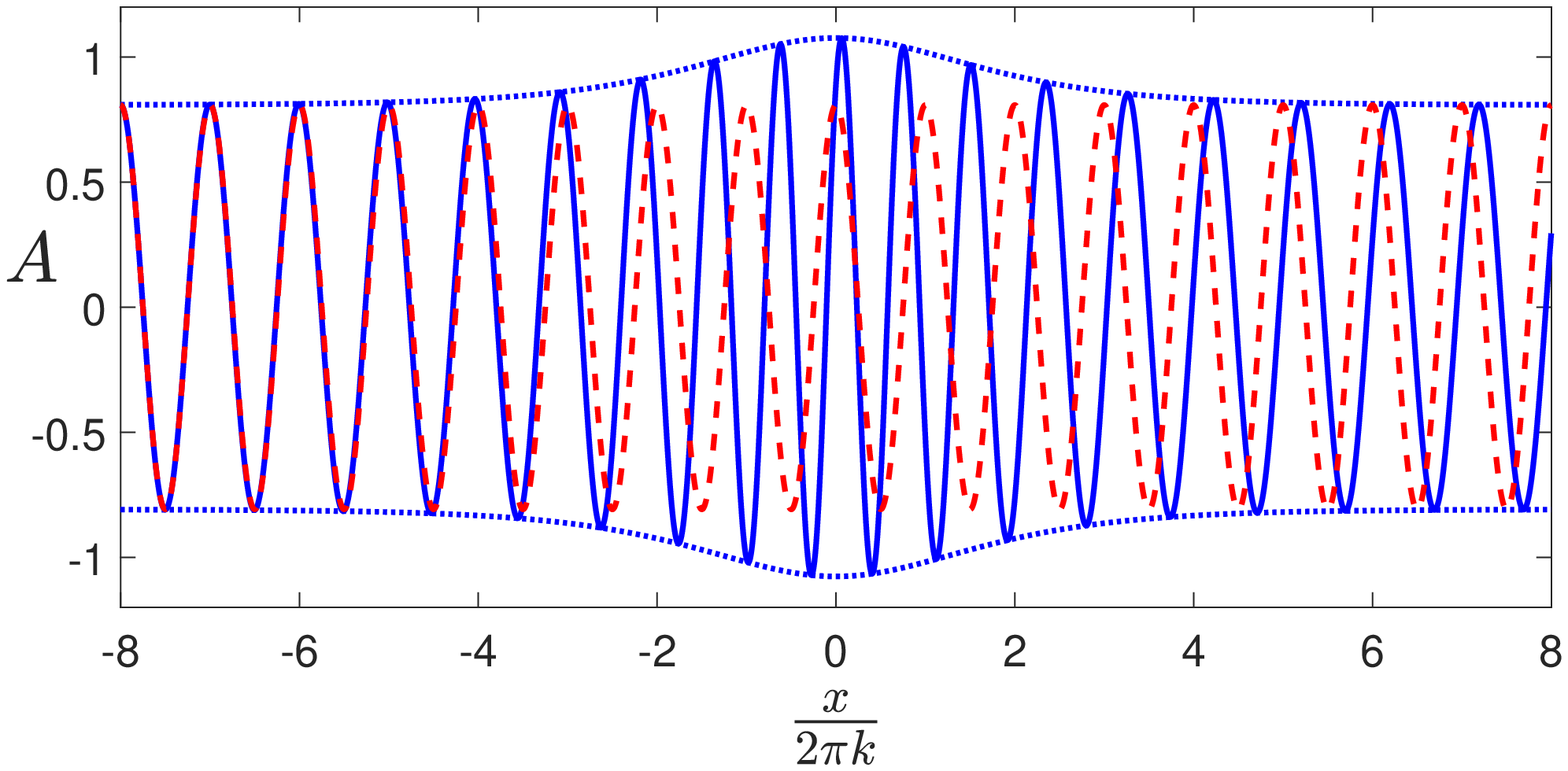}
\caption{}
\end{subfigure}
\begin{subfigure}{0.7\textwidth}
\includegraphics[width=\textwidth]{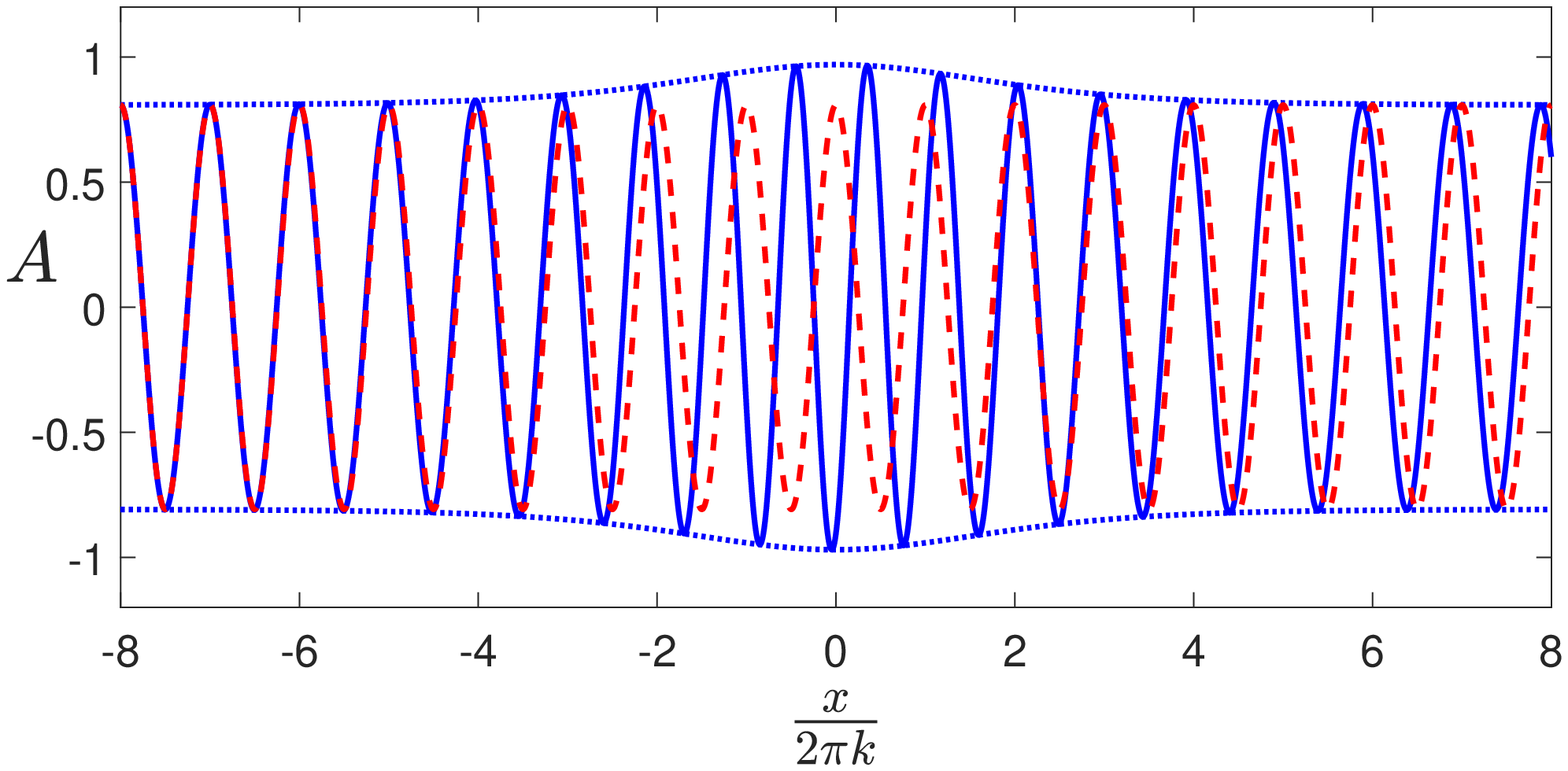}
\caption{}
\end{subfigure}
\caption{Examples of how the solution (\ref{sol-gard}) affects the plane wave solution in the regimes where (a) $B = 9.19$ and (b) $B=1.63$.}
\label{fig:gardner}
\end{figure}

The fact that solutions of both polarities are admissible in the mKdV presents the issue of how this may be selected in practice. One way to ensure a certain polarity is selected is to break this symmetry, which may be achieved by instead choosing a wavespeed close to $-2k$. We achieve this by setting
\[
c = -2k+\eps \gamma
\]
for $\gamma = \mathcal{O}(1)$. The modulation in this case instead leads to the Gardner equation
\[
\beta\Psi U_T- \frac{3\beta\Psi\gamma}{2(1+\beta k)} UU_X+\frac{3\beta^2\Psi^2}{2(1+\beta k)^2}U^2U_X+\frac{3+2\alpha^2h\Psi}{6}U_{XXX} = 0\,.
\]
The solitary waves admitted by this system differ from those arising from (\ref{mKdV}) and have more interesting forms due to the presence of $\gamma$. These solutions are given by \cite{gpt10}
\begin{equation}\label{sol-gard}
U =-\frac{4 V(1+\beta k)}{\gamma\beta \Psi\bigg[1+B \cosh\bigg(\sqrt{\frac{6\beta \Psi V}{3+2\alpha^2h\Psi}} \ \xi \bigg)\bigg]} \,, \quad B^2 = 1+\frac{4 V \beta \Psi}{\gamma^2}\,,
\end{equation}
with again the requirement that $\beta V>0$. From this, it is clear that $\gamma$ now controls the polarity of the solution, rather than both signs being admissible. There are two limits of interest for this solution. The first is as $B \rightarrow \infty$, namely as the quadratic coefficient becomes small, where the classical ${\rm sech}$ solitary wave of the mKdV (\ref{sol-mKdV}) is recovered. The other is as $B\rightarrow 1$, for which the solitary wave becomes broader. One also notes that the tabletop solitary wave solution of the Gardner equation is no longer possible, since $B^2\geq 1$. Examples of both regimes and their effect on the plane wave solution, as reconstructed according to (\ref{ansatz-mKdV}), are depicted in figure \ref{fig:gardner}. As one might expect, the regime where $B$ is large is reminiscent of the mKdV case, with more pronounced bifurcating structures than for the KdV. For lower $B$ (corresponding to larger $\gamma$) one observes narrower packets of lower amplitude, similar to those of the KdV dynamics. Thus the Gardner equation sheds light on the dynamics in the intermediate regime between the KdV and mKdV equations, and thus the way in which the solitary waves become tighter and increase in amplitude, as one may expect.

\section{Concluding Remarks}\label{sec:CR}
This paper has demonstrated that there is a great wealth of phase dynamics emerging from the Dysthe equation, primarily owing to both mean flow and self-steepening effects. The simplest of these is the KdV, which predicts how plane waves can bifurcate to modulated, or dark/bright structures. Additionally, when certain criterion are met, richer dynamics can occur by the increment of nonlinearity within the phase equations.

There are other phase reductions that may be admitted from the study of the Dysthe equation. For example when the characteristics of the Whitham equation coalesce, both the time derivative and nonlinear terms in (\ref{KdV}) vanish, and instead a modified version of the two-way Boussinesq equation arises:
\begin{equation}
U_{TT}+\bigg(\frac{\beta^2\Psi^2}{2(1+\beta k)^2}U^3+\frac{1}{6}(3-2\alpha^2 \Psi h)U_{XX}\bigg)_{XX}-\frac{\beta  \Psi}{1+\beta k}(2UU_T+U_X\partial_X^{-1}U_T)_X = 0\,.
\end{equation}
The dynamics of this equation are quite complicated, and it does not appear to admit solitary waves for the coefficients which emerge. Thus, a more delicate analysis of this equation would be necessary and would involve the study of the periodic solutions it supports. Moreover, when the Dysthe system (\ref{DystheA}) possesses higher order dispersive effects, we expect the phase dynamics to also have increased dispersive properties within certain parameter regimes. Such scenarios are expected to lead to the fifth order KdV equation emerging from the modulation, and perhaps lead to more interesting bifurcating structures.

Although the phase dynamics provides a qualitative picture as to the formation of solitary structures within the original plane wave, the next step would be to compare the results obtained here to those from direct numerical simulation. This would quantify the ability of the the nonlinear dynamics discussed here to capture the true bifurcating behaviour, and possibly help to lend these analytic techniques more credibility when discussing these scenarios.

There are other systems which exhibit a mean flow coupling within a set of nonlinear equations with a free surface. Examples include a version of (\ref{DystheA}) with higher order terms in $A$ or systems such as Benney-Roskes \cite{br69} or Hasimoto-Ono \cite{ho72} equations. These can also be explored using a phase dynamical approach to investigate how the flow beneath a wave influences how the uniform wave is modulated in the presence of defects. The novel features of the latter systems is that the wave action vectors these problems admit are nondegenerate, unlike the Dysthe equation, and so richer time dynamics are possible.

\section*{Acknowledgements}
The author would like to thank helpful discussions and feedback from Karsten Trulson, Miguel Onorato and Ying Huang during the formulation and writing of this paper.

\bibliographystyle{amsplain}

%\bibliography{Dysthe}

\appendix
{\bf \Large Appendix}

\section{Phase Dynamical Reduction to the KdV via Modulation}\label{App:Mod}
Here, we provide some details as to how the KdV equation (\ref{KdV}) may be obtained by modifying the phase modulation approach of Bridges and Ratliff \cite{br18}. Due to the similarity of the calculations, we only provide details on the key differences that arise and refer the reader to the above article for the remainder. We also note that the subsequent notation will be adopted from this work, but is consistent with that of the current paper.

The starting point is the multisymplectic form of the Lagrangian, obtained via a sequence of Legendre transformations, and has the generic structure
\begin{equation}\label{MSS}
\mathscr{L} = \iint \frac{1}{2}\langle Z,{\bf M}Z_t\rangle+\frac{1}{2}\langle Z,{\bf J}Z_x\rangle - S(Z)\,dxdt\,,
\end{equation}
for a new state variable $Z$, which contains the original state variables along with the new conjugate variables which emerge from the Legendre transforms.
The matrices ${\bf M},\,{\bf J}$ are constant skew symmetric matrices for the purposes of this analysis, however the more general problem where these depend on $Z$ follows a similar format to the reduction undertaken here, with the details on this case requiring a combination of \cite{br18} and \cite{r19a}.
The abstract set-up for the problem proceeds identically to the aforementioned work of \cite{br18}, such as the theory for the conservation laws and characteristics, Jordan chain theory and Hermitian matrix pencils. Primarily, we highlight the assumption of a two-phased relative equilibrium $\Zh(\bth;\bk,\bw)$, with
\[
\bth = \begin{pmatrix}
\theta_1\\
\theta_2
\end{pmatrix} = \begin{pmatrix}
k_1x+\omega_1t+\theta_0^{(1)}\\
k_2x+\omega_2t+\theta_0^{(2)}
\end{pmatrix}\,, \quad \bk = \begin{pmatrix}
k_1\\
k_2
\end{pmatrix}\,, \qquad \bw = \begin{pmatrix}
\omega_1\\
\omega_2
\end{pmatrix}\,.
\]
Averaging the Lagrangian over the phases of the relative equilibrium and differentiating with respect to the parameters $k_i,\,\omega_i$ results in the components of the conservation of wave action associated with each of the phases $\theta_i$,
\begin{equation}\label{cons-laws-app}
{\bf A} = 
\begin{pmatrix}
\widehat{\mathscr{L}}_{\omega_1}\\
\widehat{\mathscr{L}}_{\omega_2}
\end{pmatrix} = \frac{1}{2}
\begin{pmatrix}
\lth Z,{\bf M}Z_{\theta_1}\rth\\
\lth Z,{\bf M}Z_{\theta_2}\rth
\end{pmatrix}\,, \quad 
{\bf B} = 
\begin{pmatrix}
\widehat{\mathscr{L}}_{k_1}\\
\widehat{\mathscr{L}}_{k_2}
\end{pmatrix} = \frac{1}{2}
\begin{pmatrix}
\lth Z,{\bf J}Z_{\theta_1}\rth\\
\lth Z,{\bf J}Z_{\theta_2}\rth
\end{pmatrix}\,,
\end{equation}
where $\widehat{\mathscr{L}}$ denotes the averaged Lagrangian and  $\lth\cdot,\,\cdot\rth$ is a suitable $\theta_i-$averaging inner product. These would be the vector-valued functions which satisfy the multiphase version of the Whitham equations (\ref{WME}) at the scales $X = \eps x,\, T = \eps t$, but their derivatives will be used to form the coefficients within the KdV obtained here under a different scaling of the slow variables detailed below.  We also note that within the reduction procedure, the solvability requirement for the inhomogenous problems involving the linear operator
\[
{\bf L} = \D^2S(\Zh)-\sum_{i=1}^2\big( \omega_i{\bf M}+k_i{\bf J}\big)\partial_{\theta_i}\,,
\]
which emerge is precisely
\[
{\bf L}F = G \quad \mbox{is solvable if and only if} \quad \lth \Zh_{\theta_1},G\rth = 0 = \lth \Zh_{\theta_2},G\rth = 0\,.
\]
This is since the kernel of the linear operator ${\bf L}$ is assumed to be the span of $\Zh_{\theta_1},\,\Zh_{\theta_2}$. More details on all of these concepts can be found in the earlier work of \cite{br18}.

The first departure from the theory of the work of \cite{br18} is to instead use the ansatz
\begin{multline}\notag
Z = \Zh\big(\bth+\eps \bph,\,\bk+\eps^2 \bq,\,\bw+\eps^2c\bq+\eps^4 \bO\big)\\
+\eps^3\big( W_0(\bth,X,T)+\eps W_1(\bth,X,T)+\eps^2W_2(\bth,X,T)\big)\,,
\end{multline}
with
\[ 
X = \eps (x+ct)\,,\qquad T = \eps^3 t\,,
\]
and
\[
\bph(X,T) = \begin{pmatrix}
\phi_1(X,T)\\
\phi_2(X,T)
\end{pmatrix}\,, \qquad
\bq = \begin{pmatrix}
q_1(X,T)\\
q_2(X,T)
\end{pmatrix} = \bph_X\,, \qquad \bO = 
\begin{pmatrix}
\Omega_1(X,T)\\
\Omega_2(X,T)
\end{pmatrix} = \bph_T\,.
\]
This is then substituted into the Euler-Lagrange equations associated with (\ref{MSS}), a Taylor expansion around the state $\eps = 0$ is undertaken and the resulting sequence of equations are solved for each order of $\eps$. The leading, first, second and third orders result in exactly the same systems to solve as \cite{br18}, and the first two of these are automatically satisfied by properties of the solution $\Zh$. At third order, we recover
\begin{equation}\label{third-ord-app}
{\bf L}W_0 = {\bf K}\sum_{i=1}^2(q_i)_X\big(\Zh_{k_i}+c\Zh_{\omega_i}\big)\,, \qquad {\bf K} = {\bf J}+c{\bf M}\,,
\end{equation}
and we note that the above equation can be shown to be solvable precisely when
\begin{equation}\label{Del-c-app}
\Delta(c) = {\rm det}\big[c^2 \D_\bw{\bf A}+c(\D_\bk{\bf A}+\D_\bw{\bf B})+\D_\bk{\bf B}\big] \equiv {\rm det}\big[{\bf E}(c) \big] = 0\,.
\end{equation}
This is exactly the condition for $c$ to be a characteristic of the Whitham modulation equations. The key assumption in this analysis is that all of the roots are distinct, that is we are assuming  that $\Delta\red{'}(c)\neq 0$ for any of the $c$ which satisfy the above. Overall, this vanishing determinant imposes that
\[
\bq = U(X,T)\be\,, \quad \mbox{where} \quad {\bf E}(c)\begin{pmatrix} \zeta_1\\\zeta_2\end{pmatrix}\equiv  {\bf E}(c)\be = {\bf 0}\,,
\]
namely that $\be$ is the eigenvector associated with the zero eigenvalue of ${\bf E}$.
This leads to the expression for $W_0$ in (\ref{third-ord-app}) as
\[
W_0 = U_X{\bf v}_3+\sum_{i=1}^2\alpha_i\Zh_{\theta_i}\,, \quad \mbox{where} \quad {\bf L}{\bf v}_3  =  {\bf K}\sum_{i=1}^2\zeta_i\big(\Zh_{k_i}+c{\bf M}\Zh_{\omega_i}\big)\,.
\]
The functions $\alpha_i(X,T)$ are required to prevent the equation emerging from the modulation reduction to simply admit the trivial solution. Subsequently, the fourth order in $\eps$ is simpler than that appearing in \cite{br18}, and is simply
\[
{\bf L}\bigg[W_1-\sum_{i=1}^2\big(\zeta_i\phi_iU_{X}({\bf v}_3)_{\theta_i}+(\alpha_i)_X(\Zh_{k_i}+c\Zh_{\omega_i})\big)\bigg] = U_{XX}{\bf K}{\bf v}_3\,.
\]
The right hand side of this expression lies in the range of the linear operator ${\bf L}$, as the zero eigenvalue of ${\bf L}$ is of even algebraic multiplicity. Therefore, the solution at this order is simply
\[
W_1 = U_{XX}{\bf v}_4+\sum_{i=1}^2\zeta_i\phi_iU_{X}({\bf v}_3)_{\theta_i}\,, \quad \mbox{where} \quad {\bf L}{\bf v}_4 = {\bf K}{\bf v}_3\,.
\]

The final, and crucial, order of the modulation analysis at which the KdV equation emerges leads to the equation
\[
\begin{split}
{\bf L}\widetilde{W}_2 =& \ U_T\sum_{i=1}^2\zeta_i\big({\bf J}\Zh_{\omega _i}+{\bf M}\Zh_{k_i}+2c{\bf M}\Zh_{\omega_i}\big)\\
 &+ UU_X\sum_{i=1}^2\bigg[{\bf K}({\bf v}_3)_{\theta_i}-\D^3S(\Zh)({\bf v}_3,\Zh_{k_i}+c\Zh_{\omega_i})\\[2mm]
&\hspace{2.5cm}\displaystyle +\sum_{j=1}^2{\bf K}(\Zh_{k_ik_j}+c\big(\Zh_{\omega_ik_j}+\Zh_{k_i\omega_j}\big)+c^2\Zh_{\omega_i\omega_j})\bigg]\\
&+U_{XXX}{\bf K}{\bf v}_4+{\bf K}\sum_{i=1}^2(\alpha_i)_{XX}(\Zh_{k_i}+c\Zh_{\omega_i})\,.
\end{split}
\]
The term $\widetilde{W}_2$ is defined as the sum of $W_2$ with all terms at this order which are pre-images of the expressions which lie in the range of ${\bf L}$ on the right hand side. Its form is not important, since an exact expression for this is only required if the analysis proceeds to further orders in $\eps$, however it terminates here. All that remains is to take the inner product of the right hand side of the above with the kernel elements $\Zh_{\theta_1},\,\Zh_{\theta_2}$ and set this to zero, thus imposing that the right hand side also lies in the range of ${\bf L}$, which generates the KdV equation which $U$ must satisfy. In fact, all of the coefficients of the relevant terms have already been computed in \cite{br18}, and so we simply state the results of these inner products. Firstly, the inner product of the terms involving $U_T$ lead to the vector
\[
\begin{split}
\begin{pmatrix}
\lth \Zh_{\theta_1},\sum_{i=1}^2\zeta_i\big({\bf J}\Zh_{\omega _i}+{\bf M}\Zh_{k_i}+2c{\bf M}\Zh_{\omega_i}\big)\rth\\
\lth \Zh_{\theta_2},\sum_{i=1}^2\zeta_i\big({\bf J}\Zh_{\omega _i}+{\bf M}\Zh_{k_i}+2c{\bf M}\Zh_{\omega_i}\big)\rth
\end{pmatrix}U_T &= -\big(2c\D_\bw{\bf A}+\D_\bk{\bf A}+\D_\bw{\bf B} \big)\be U_T\\
&\equiv -{\bf E}'(c)\be U_T\,,
\end{split}
\]
as is found as part of equation (5.12) in \cite{br18}. Next, the terms involving the quadratic term $UU_X$ give the vector
\[
\begin{split}
-(\D_\bk^2{\bf B}\red{+}c(2\D_\bk\D_\bw{\bf B}+\D_\bk^2{\bf A})+c^2(2\D_\bk\D_\bw{\bf A}+\D_\bw^2{\bf B})\red{+}c^3\D_\bw^2{\bf A})(\be,\be)UU_X\\
\equiv -{\bf H}(\be,\be) UU_X\,,
\end{split}
\]
when the inner product is taken, which can be found in equation (5.21) of \cite{br18}. Finally, the terms which involve $U_{XXX}$ lead to the vector
\[
\begin{pmatrix}
\lth \Zh_{\theta_1},({\bf J}+c{\bf M}){\bf v}_4\rth\\
\lth \Zh_{\theta_2},({\bf J}+c{\bf M}){\bf v}_4\rth
\end{pmatrix}U_{XXX} = -{\bf T}U_{XXX}\,,
\]
which is a vector involving nonzero constants arising from the termination of the relevant Jordan chain. The $\alpha_i$ terms simply give $-{\bf E}(c){\bm \alpha}_{XX}$ with ${\bm \alpha} = (\alpha_1,\,\alpha_2)^T$. Therefore, we arrive at the vector equation
\[
{\bf E}'(c)\be U_T+{\bf H}(\be,\be) UU_X+{\bf T}U_{XXX}+{\bf E}(c){\bm \alpha}_{XX} = 0\,,
\]
which may then be projected to a scalar equation by multiplying on the left by $\be$ (which eliminates the ${\bm \alpha}$ term), giving
\[
\be^T{\bf E}'(c)\be U_T+\be^T{\bf H}(\be,\be) UU_X+\be^T{\bf T}U_{XXX} = 0\,.
\]
Thus the KdV equation emerges from the modulation with the characteristic moving frame, with only slight modifications of the existing theory. The key assumption that $\Delta'(c) \neq 0$ ensures that the coefficient of the $U_T$ term above is nonzero, but this will vanish whenever there is a repeated root of $\Delta(c)$, in which case one returns exactly to the modulation analysis in \cite{br18}.

\section{Phase Dynamical Reduction to the KdV via the Madelung Transform}\label{App:PD}
Here, we provide details as to how the KdV equation can be derived from the Dysthe equation (\ref{DystheA}) without the explicit use of a modulation argument. To do so, we first undertake a Madelung transform of the Dysthe equation by introducing
\[
A = \sqrt{\rho(x,t)}e^{i\phi(x,t)}\,,
\]
for real functions $\rho,\,\phi$, and splitting the resulting system in $(\rho,\phi,\Phi)$ into real and imaginary parts. This leads to the system of equations
\begin{equation}\label{DystheMad}
\begin{split}
\frac{1}{2}\rho_t+\bigg(\rho u+\frac{\beta}{4}\rho^2\bigg)_x &= 0\,,\\[2mm]
u_t+2uu_x+\beta(\rho u)_x+\rho_x-\alpha \left.\Phi_{xx}\right\vert_{z=0} &= \bigg(\frac{\rho_{xx}}{2\rho}-\frac{\rho_x^2}{\rho^2}\bigg)_x\,,\\[2mm]
\left.\Phi_z\right\vert_{z=0} &= \alpha \rho_x\,,\\[2mm]
\Phi_{xx}+\Phi_{zz} &= 0\,, \quad z \in (-h,0)\,,\\[2mm]
\left.\Phi_z\right\vert_{z=-h} &= 0\,,
\end{split}
\end{equation}
with $u = \phi_x$. One may then undertake a typical multiple scales analysis of the form
\[
\begin{split}
\rho& = \Psi+\eps^2H(X,T)+\eps^4 G(X,T)\,, \\[2mm]
 u &= k+\eps^2V(X,T)+\eps^4W(X,T)\,,\\[2mm]
 \Phi &= u_0x+\eps \Gamma(X,T)+\eps^3\bigg(\Upsilon_1(X,T)-\frac{1}{2}\Upsilon_2(X,T)(z+h)^2\bigg)\\
 &\quad+\eps^5\bigg(\Psi_1(X,T)-\frac{1}{2}\Psi_2(X,T)(z+h)^2+\frac{1}{24}\Psi_3(X,T)(z+h)^4\bigg)\,,
\end{split}
\]
with $X = \eps (x+ct),\,T = \eps^3t$. The basic state\red{s} for $\rho,\,u$ and $\Phi$ have been chosen so that they match those given in \S \ref{sec:KdV}. The leading, first and second orders in $\eps$ are automatically satisfied by the above expansion, whereas the third order in $\eps$ leads to the linear system
\[
{\bf F}\begin{pmatrix}
H\\
V\\
\Gamma_X
\end{pmatrix}_X = \begin{pmatrix}
\frac{1}{2}(2k+\beta \Psi+c)&\Psi&0\\
1+\beta k&2k+\beta \Psi+c&-\alpha\\
\alpha&0&h
\end{pmatrix}
\begin{pmatrix}
H\\
V\\
\Gamma_X
\end{pmatrix}_X  = {\bf 0}\,.
\]
This system is therefore solvable whenever the determinant of the above matrix vanishes, giving that
\[
\frac{h}{2}(2k+\beta \Psi+c)^2-\Psi\big(h(\beta k+1)+\alpha^2\big) = 0\,,
\]
recovering the criticality condition (\ref{c-defn}). This gives that $H,\,U$ and $\Gamma_X$ should all be related via the eigenvector of the zero eigenvalue of ${\bf F}$,
\begin{equation}\label{HVG-rel}
\begin{pmatrix}
H\\V\\\Gamma_X
\end{pmatrix} = \begin{pmatrix}
-2\Psi\\
2k+\beta\Psi+c\\
\frac{2\Psi \alpha}{h}
\end{pmatrix}U\,,
\end{equation}
and we will also require the left eigenvector
\[
\eta = \bigg(-\frac{2k+\beta \Psi+c}{\Psi},1,\frac{\alpha}{h}\bigg)^T\,, \quad \mbox{where} \quad {\bf F}^T\eta = {\bf 0}\,.
\]
At the final order, $\eps^5$, the system can be written as
\[
{\bf F} \begin{pmatrix}
G\\W\\ (\Upsilon_1)_X
\end{pmatrix}_X = - \begin{pmatrix}
\frac{1}{2}H_T+\bigg(HV+\frac{\beta}{4}H^2\bigg)\\
V_T+2VV_X+\beta(HV)_X-\frac{1}{2\Psi}H_{XXX}+\frac{\alpha h^2}{2}\Gamma_{XXXX}\\
-\frac{h^3}{6}\Gamma_{XXXX}
\end{pmatrix}\,.
\]
For the right hand side of this system to lie in the range of ${\bf F}$, it must vanish when multiplied on the left by $\eta$. Doing so, and replacing $H,\,V$ and $\Gamma$ according to (\ref{HVG-rel}) gives
\[
(2k+\beta \Psi+c)(U_T+3(2k+c)UU_X)+\frac{1}{6}(3+2\alpha^2\Psi h)U_{XXX} = 0\,,
\]
giving the same KdV equation as (\ref{KdV}) up to scaling, namely by redefining
\[
U \mapsto \frac{\big(h(\beta^2\Psi+\beta c-2)-\alpha^2\big)}{2(2k+c)(1+\beta k)}U\,.
\]
This factor is nonsingular for the cases considered within this paper, and in particular we note that in the criticality that leads to the mKdV discussed within this paper this factor becomes $\frac{1}{2(1+\beta k)}$ as the $2k+c$ factor in the denominator cancels with the numerator in this scenario.

A similar approach can be used to obtain the remaining reductions discussed within this paper, but requires a different ansatz, which we provide below but without the details of the reduction. This is because the analyses are very similar to that of the KdV. For the modified KdV equation, one must use
\[
\begin{split}
\rho& = \Psi+\eps H(X,T)+\eps^2 G(X,T)+\eps^3F(X,T)\,, \\[2mm]
 u &= k+\eps V(X,T)+\eps^2W(X,T)+\eps^3R(X,T)\,,\\[2mm]
 \Phi &= u_0x+ \Gamma(X,T)+\eps^2\bigg(\Upsilon_1(X,T)-\frac{1}{2}\Upsilon_2(X,T)(z+h)^2\bigg)\\
 &\quad+\eps^3\bigg(\Psi_1(X,T)-\frac{1}{2}\Psi_2(X,T)(z+h)^2+\frac{1}{24}\Psi_3(X,T)(z+h)^4\bigg)\\
  &\quad+\eps^4\bigg(\Xi_1(X,T)-\frac{1}{2}\Xi_2(X,T)(z+h)^2\\
  & \hspace{3.5cm}+\frac{1}{4!}\Xi_3(X,T)(z+h)^4-\frac{1}{6!}\Xi_4(X,T)(z+h)^6\bigg)\,,
\end{split}
\]
again taking $X = \eps (x+ct),\,T = \eps^3t$ and assuming (\ref{mKdV-cond}) holds. For the modified two-way Boussinesq one uses the above ansatz but with $T = \eps^2t$ and instead assuming that $\Delta (c)$ in (\ref{c-cond}) leads to a double root.

\end{document}